\documentclass[aps,prd,floatfix,nofootinbib,superscriptaddress,twocolumn,showkeys]{revtex4-2}

\usepackage{amsmath}%,amssymb}

\usepackage{amssymb}
\allowdisplaybreaks

\usepackage{makeidx}
\usepackage{amsfonts}
\usepackage{dutchcal}
\usepackage[usenames,dvipsnames]{pstricks}
\usepackage{subfigure}
\usepackage{epsfig}
\usepackage{pst-grad} % For gradients
\usepackage{mathtools}
\usepackage[colorlinks,hyperindex]{hyperref}
\usepackage{array}
\hypersetup
{	colorlinks,%
	citecolor=black,%
	linkcolor=black,%
	urlcolor=black,%
}

\usepackage{allpurpose}

\usepackage{flushend}% Align the end of the last page.
\newcommand\nn{{\nonumber}}

\begin{document}

\title{Deflection and Gravitational lensing of null and timelike signals in the Kiselev black hole spacetime in the weak field limit}

\author{Haotian Liu}
\thanks{These authors contributed equally to this work.}
\address{School of Physics and Technology, Wuhan University, Wuhan, 430072, China}

\author{Jinning Liang}
\thanks{These authors contributed equally to this work.}
\address{School of Physics and Technology, Wuhan University, Wuhan, 430072, China}

\author{Junji Jia}
\email[Corresponding author:~]{junjijia@whu.edu.cn}
\address{MOE Key Laboratory of Artificial Micro- and Nano-structures, School of Physics and Technology, Wuhan University, Wuhan, 430072, China}

\date{\today}

\begin{abstract}
In this work we study the deflection and gravitational lensing of null and timelike signals in the Kiselev spacetime in the weak field limit, to investigate the effects of the equation of state parameter $\omega$ and the matter amount parameter $\alpha$. In doing this, we extend a perturbative method  previously developed for asymptotically flat spacetimes whose metric functions have integer-power asymptotic expansions to the case that may or may not be asymptotically flat but with non-integer power expansions. 
It is found that in the asymptotically flat case ($-1/3<\omega<0$) the deflection angles are expressable as quasi-power series of the dimensionless quantities $M/b,~b/r_{s,d}$ and $\alpha/M^{1+3\omega}$ where $M,~b,~r_{s,d}$ are respectively the lens mass, impact parameter and source/detector radius. A similar series exists for the non-asymptotically flat case of ($-1<\omega<-1/3$), but with the closest radius $r_0$ replacing $b$. In the asymptotically flat (or non-flat) case, the increase of $\alpha$ or decrease of $\omega$ will increase (or increase) the deflection angle.   
Since the obtained deflection angles naturally take into account the finite distance effect of the source and the detector, we can establish an exact gravitational lensing equation, from which the apparent angles of the images and their magnifications are solved. It is found that generally for the asymptotically flat case, increasing $\alpha$ or decreasing $\omega$ will increase the apparent angles of the images. While for the non-asymptotically flat case, increasing $\alpha$ or $\omega$ will both lead to smaller apparent angles.
\end{abstract}


\maketitle

\section{Introduction}

Deflection of light and gravitational lensing (GL) have played very important roles in the acceptance of General Relativity as a correct description of gravity \cite{Dyson:1920cwa}, and in the development of astronomy \cite{Walsh:1979nx} (see \cite{Bartelmann:1999yn, Perlick:2004tq} for reviews on GL). Nowadays, GL is used very broadly in both theoretical and observational studies involving gravity. 
It has been used to measure the Hubble constant \cite{Refsdal:1964nw,Kundic:1996tr}, to map (super)cluster and their group's mass \cite{Gray:2001zx,Hoekstra:2000ux}, and to probe dark energy and new gravity theories \cite{Hoekstra:2008db,Joyce:2016vqv}, just to name a few. 

Traditionally messengers in such deflection and GL are only light signals. Due to the observation of neutrinos from extragalactic sources \cite{Hirata:1987hu, Bionta:1987qt,IceCube:2018dnn,IceCube:2018cha}, gravitational waves \cite{Abbott:2016blz,Abbott:2016nmj,TheLIGOScientific:2017qsa} and GL of supernovae \cite{Kelly:2014mwa,Goobar:2016uuf} in recent years, as well as the more historical cosmic rays \cite{LetessierSelvon:2011dy}, it is clear that in principle timelike signals can also experience deflection and act as messengers in GL. From the field strength viewpoint, then the traditional GLs are all based on the deflection in the weak field limit, i.e., the deflection angles are small. However, with the more recent successful observation of the M87$^{*}$  \cite{EventHorizonTelescope:2019dse,EventHorizonTelescope:2019ggy} and Sgr A$^{*}$ \cite{EventHorizonTelescope:2022xnr,EventHorizonTelescope:2022xqj} supermassive black hole (SMBH) images by the Event Horizon Telescope (EHT) collaboration, we have gained the ability to observe signals experienced deflections in a strong field. Therefore, theoretical investigation of these unconventional messengers' deflection and GL in both weak and strong field limits has become more popular lately using both perturbative method \cite{Jia:2020xbc,Huang:2020trl} and Gauss-Bonnet theorem method \cite{Gibbons:2008rj,Crisnejo:2018uyn,Li:2019qyb}.

On the other hand, in recent years, to explain the observed accelerated expansion of the Universe, an enormous research effort has been made on building various models of gravity and dark energy \cite{Frieman:2008sn}. Among these, the quintessence model is uniquely appealing because of its connections with dynamical fields/potentials, many of which naturally arise from particle physics \cite{Caldwell:2005tm, Tsujikawa:2013fta}. Although quintessence is mostly used in cosmological models, there has also been great interest in seeking static and spherically or axially symmetric spacetime 
models possessing a ``quintessence'' \cite{Kiselev:2002dx,Toshmatov:2015npp} and studying their properties. One of the models that are intensively studied  is obtained in Ref \cite{Kiselev:2002dx}. It was initially introduced as a ``quintessence'' model due to the equation of state (EOS) $p=\omega\rho$ for the matter involved in this spacetime. However later on, it was shown that this model is not really a quintessence one in the conventional sense \cite{Visser:2019brz}. In what follows, therefore we will refer to the spacetime in Ref. \cite{Kiselev:2002dx} as the Kiselev BH spacetime. We emphasize that the matter in this spacetime does allow a variable EOS parameter $\omega$ so that it can mimic several familiar spacetimes.

A more practical reason that we investigate the Kiselev spacetime comes from the recent progress in the observation of the M87$^{*}$ and Sgr A$^{*}$ SMBHs \cite{EventHorizonTelescope:2019dse,EventHorizonTelescope:2019ggy,GRAVITY:2021xju,EventHorizonTelescope:2022xnr,EventHorizonTelescope:2022xqj}. Although black holes in many astrophysical studies are assumed by default to be the Kerr type, excluding other possibilities however are not very easy. Even the EHT team themselves have studied the possibility that these SMBHs are of other types \cite{EventHorizonTelescope:2022xqj}, and there are already works assuming them to be black holes in the Kiselev spacetime \cite{Xu:2018mkl,Das:2021otl,Abbas:2021whh}. In principle, the signals forming these shadows mainly originate from the innermost stable circular orbit of the accretion disk and they might have circled around the photon sphere before reaching us. Even though these regions are in the strong field limit of gravity while our paper is in the weak field limit, as we will see however the mass data in these observations can restrict the parameter space of $(\alpha,~\omega)$ that we will study for the Kiselev spacetime. More importantly, as we will show in this work, the observables in GL in this spacetime, including the apparent angles and time delays of the images,  are generally sensitive to the parameters $(\alpha,~\omega)$ and therefore can be used to constrain their values in the future. 

In this work, we would like to study how in general such matter characterized by $(\alpha,~\omega)$ would influence the deflection and GL of both null and timelike signals in the weak field limit in the Kiselev spacetime. 
Previously, some authors considered the deflection of light in the weak field limit in this spacetime only for specific values of $\omega$. Malakolkalami and K. Ghaderi \cite{Malakolkalami:2015tsa}, Fernando \cite{Fernando:2012ue} and Younas et al. \cite{Younas:2015sva} considered only the special case of $\omega=-2/3$. Shchigolev and  Bezbatko considered the case $\omega=-1/3$ and $\omega=-2/3$ using the homotopy-perturbation method \cite{Shchigolev:2016gro}. 
He and Zhang considered the deflection using a post-Newtonian and effective reflective index approach \cite{He:2017alg}. Azreg-A\"{i}nou et al. considered the deflection of light in charged Kiselev BH \cite{Azreg-Ainou:2017obt}.
Others considered even simpler choices such as $\omega=0$ for the Schwarzschild case, $\omega=1/3$ for the Reissner-Nordstr\"{o}m (RN) case and $\omega=-1$ for Schwarzschild-de Sitter (SdS) case \cite{Ghaderi:2017wvl,Zhang:2021ygh}. 

All the above works considered only the deflection of null rays and most of them worked with infinite source and observer radii.
In contrast, our consideration has the following advantages. First, it is applicable to the deflection of both null and timelike rays with {\it arbitrary} EOS parameter $\omega$. Moreover, our method takes into account the finite distance effect of the source and observer to the deflection angle naturally, and the resultant deflection angle allows us to use an exact lensing equation. Lastly, these features allow us to study the effects of $\omega$ and $\alpha$ on the apparent angles and magnifications in this spacetime, which were seldom considered before. Throughout this work, we use geometrized units $(G=c=1)$ 
and metric signature $(-,+,+,+)$.

\section{The perturbative method}

We start from the general static spherically symmetric metric
\begin{equation}
\dd s^2=-A(r) \dd t^2+B(r)\dd r^2+C(r) (\dd\theta^2+\sin^2\theta\dd\phi^2). \label{eq:sssmetric}
\end{equation}
where $(t,~r,~\theta,~\phi)$ are the coordinates and $A(r),~B(r)$ and $C(r)$ are the metric functions. Although locally we can always choose $C(r)=r^2$, for now we will keep the general form of $C(r)$ since some metrics are written in a different coordinate system.

To compute the deflection angle, we will use the perturbative method developed in Ref. \cite{Huang:2020trl, Liu:2020mkf}. In the following, we will briefly recap the method and apply the procedure directly to the Kiselev BH spacetime in Sec.  \ref{ssec:apptoqs}.
For the spacetime described by \eqref{eq:sssmetric}, the geodesic equations read
\begin{align}
&\dot{t}=\frac{E}{A},\label{eq:teq}\\
&\dot{\phi}=\frac{L}{C}, \label{eq:phieq}\\
&\dot{r}^2=\frac{(E^2-\kappa A)C-L^2A}{ABC},\label{eq:req}
\end{align}
where $\kappa=0,~1$ for null and timelike signals respectively, dot means derivative with respect to the proper time or affine parameter $\lambda$. Without losing any generality we have set the trajectory to be in the equatorial plane, i.e., $\theta=\pi/2$. The $L$ and $E$ here are the constants of the first integrals. We can define an effective potential $V_{\mathrm{eff}}(r)$ in the right hand side of Eq. \eqref{eq:req}
\be 
V_{\mathrm{eff}}(r)=\frac{L^2}{2BC}+\frac{\kappa}{2B} \label{eq:gveffdef}
\ee 
to help us to understand the behavior of the deflection angles in Sec. \ref{ssec:apptoqs}. With this, Eq. \eqref{eq:req} becomes
\be 
\frac{\dot{r}^2}{2}=\frac{E^2}{2AB}-V_{\mathrm{eff}}(r).\label{eq:reqmod}
\ee

Using Eqs. \eqref{eq:teq} to \eqref{eq:req}, the corresponding change of the angular coordinate $\Delta\phi$ and total flight time of signal from the source at $(r_s,\theta_s)$ to the detector at $(r_d,\phi_d)$ become respectively (see Fig. \ref{fig:gl})
\be
\Delta\phi=\lsb \int_{r_0}^{r_s}+\int_{r_0}^{r_d}\rsb \sqrt{\frac{B}{C}}\frac{L}{\sqrt{(E^2/A-\kappa )C-L^2}}\dd r.
\label{eq:deltaphi}
\ee
and
\be
\Delta t=\lsb \int_{r_0}^{r_s}+\int_{r_0}^{r_d}\rsb \frac{E}{A}\frac{\sqrt{BC}}{\sqrt{(E^2/A-\kappa )C-L^2}}\dd r
\label{eq:totaltime}
\ee
Here $r_0$ is the minimal radial coordinate of the trajectory. It can be related to $L$ using Eq. \eqref{eq:req}, i.e. $\dot{r}|_{r=r_0}=0$ to find
\be
L=\sqrt{[E^2-\kappa A(r_0)]C(r_0)/A(r_0)}. \label{eq:linr0}
\ee
The integrals in Eqs. \eqref{eq:deltaphi} and \eqref{eq:totaltime} usually can not be explicitly carried out for general metrics and therefore require a perturbative technique to find an approximation.

In an asymptotically flat spacetime, $L$ and $E$ can be interpreted respectively as the angular momentum and energy per unit mass of the timelike signal. They can also be related to the signal velocity $v$ at infinity and the impact parameter $b$ using
\be
L=|\vecr\times \vp|=\frac{v}{\sqrt{1-v^2}}b,~E=\frac{1}{\sqrt{1-v^2}}, \label{eq:leandbv}
\ee
so that 
\be \frac{L}{E}=bv. \ee
This last equation holds for null signals too. If the spacetime is not asymptotically flat as in the SdS case, then although we can still define an effective impact parameter $b$ through the same Eq. \eqref{eq:leandbv}, its geometrical meaning as the distance between the asymptotic line and the parallel line through the center is lost for timelike signals. We will deal with the deflection angle in non-asymptotically flat spacetime in Sec. \ref{subsec:anonflat} separately and the remainder of this section is for asymptotic spacetime only.

Using Eq. \eqref{eq:leandbv}, $L$ and $E$ can always be replaced by $b$ and $v$. Further using Eq. \eqref{eq:linr0}, we can establish the following correspondence between $b$ and $r_0$
\be
\frac{1}{b}=\frac{\sqrt{E^2-\kappa}}{\sqrt{E^2-\kappa A(r_0)}}\sqrt{\frac{A(r_0)}{C(r_0)}}\equiv p\lb \frac{1}{r_0}\rb \label{eq:pfuncdef}
\ee
where in the right-hand side of the equation we defined a function $p$ of $1/r_0$. To use later, we also denote the inverse function of $p(x)$ as $q(x)$.

Now one of the difficulties in the integration of Eq. \eqref{eq:deltaphi} comes from the fact that the minimal radius $r_0$, which is usually difficult to link to observables, appears in the lower limit. Therefore in Ref. \cite{Huang:2020trl,Liu:2020mkf} we proposed the change of variables from $r$ to $u$, which are linked by
\be
\frac{1}{r}=q\lb \frac{u}{b}\rb. \label{eq:rtoucov}
\ee
More explicitly, by inverting this function and using Eq. \eqref{eq:pfuncdef}, we can also express $u$ in terms of $r$ as
\be 
u=b\cdot p\lb \frac1r\rb = \frac{b\sqrt{E^2-\kappa}}{\sqrt{E^2-\kappa A(r)}}\sqrt{\frac{A(r)}{C(r)}}.
\ee
This shows that once the metric functions are specified, then this change of variables is immediately known.
For $\Delta\phi$, substituting Eq. \eqref{eq:rtoucov} into Eq. \eqref{eq:deltaphi}, the entire integrand and the integral limits can be transformed into the following (see \cite{Huang:2020trl} for details)
\be
\Delta\phi=\lsb \int_{\sin\theta_s}^{1}+\int_{\sin\theta_d}^{1}\rsb y\lb \frac{u}{b} \rb \frac{\dd u}{\sqrt{1-u^2}} \label{eq:phiinu}
\ee
where
\begin{equation}
y\lb \frac{u}{b}\rb =\sqrt{\frac{B(1/q)}{C(1/q)}}\frac{1}{p'(q)q^2}\frac{u}{b},~~q=q\lb \frac{u}{b}\rb. \nn
\end{equation}
and
\be
\theta_{s,d}=\arcsin\lsb b\cdot p\lb \frac{1}{r_{s,d}}\rb \rsb
\label{eq:thetasdinp}
\ee
are respectively the apparent angle of the signal at the source and detector. 
For $\Delta t$, substituting Eq. \eqref{eq:rtoucov} into Eq. \eqref{eq:totaltime}, it is transformed into 
\be
\Delta t=\lsb \int_{\sin\theta_s}^{1}+\int_{\sin\theta_d}^{1}\rsb z\lb \frac{u}{b} \rb \frac{\dd u}{u\sqrt{1-u^2}} \label{eq:tinu}
\ee
where
\begin{equation}
z\lb \frac{u}{b}\rb =\frac{\sqrt{B(1/q)C(1/q)}}{A(1/q)}\frac{u}{b}\frac{1}{v}\frac{1}{p'(q)q^2}\frac{u}{b}
\end{equation}

Using Eq. \eqref{eq:pfuncdef} for $b$ and $p(1/r_{s,d})$, these apparent angles can also be recast into
\be
\theta_{s,d} = \arcsin \lb \sqrt{\frac{E^2-\kappa A(r_0)}{E^2-\kappa A(r_{s,d})}}\sqrt{\frac{A(r_{s,d})C(r_0)}{A(r_0)C(r_d)}} \rb.\label{eq:agform2}
\ee
Now we restrict ourselves to the case of the weak field limit, in which the impact parameter is much larger than the characteristic mass of the spacetime. Therefore in this limit we can expand the $y\lb \frac{u}{b}\rb $ and $z\lb \frac{u}{b}\rb$ factors in Eqs. \eqref{eq:phiinu} and \ref{eq:tinu} into power series of $\frac{u}{b}$, i.e.,
\begin{align}
&y\lb \frac{u}{b}\rb =\sum_{n\in S} y_n \lb  \frac{u}{b}\rb^n, \label{eq:yexp}\\
& z\lb \frac{u}{b}\rb =\sum_{n\in F} z_n \lb  \frac{u}{b}\rb^n, \label{eq:zexp}
\end{align}
where $S$ and $F$ are the sets of powers which does not necessarily contain only integers.
Substituting into Eqs. \eqref{eq:phiinu} and \eqref{eq:tinu}, $\Delta\phi$ and $\Delta t$ become sums of series of integrals whose $u$ dependent part takes a simple form of $\displaystyle \frac{u^n}{\sqrt{1-u^2}}$. These integrals over $u$ then can always be carried out easily, so that
\be
 \int_{\sin\theta_i}^{1}
\frac{u^n}{\sqrt{1-u^2}}\dd u \equiv I_n(\theta_i),~~~~(i=s,~d)
\ee
whose explicit forms are given in Eq. \eqref{eq:inthetares} in Appendix \ref{sec:appd}.
Overall therefore, we found an effective way to approximate $\Delta\phi$ and $\Delta t$
\begin{align}
&\Delta\phi =\sum_{i=s,d}\sum_{n=0}^\infty y_n\frac{I_n(\theta_i)}{b^n} . \label{eq:dphifinal}\\
&  \Delta t =\sum_{i=s,d}\sum_{n=0}^\infty z_n\frac{I_{n-1}(\theta_i)}{b^n} . \label{eq:dtfinal}\end{align}

A few comments are in order here.
The first is that when inverting the function $p(x)$,  an explicit and closed form of $q(x)$ might not be possible although $p(x)$ can always be known explicitly from the metric function. Fortunately, however, what is needed in Eqs. \refer{eq:yexp} and \eqref{eq:zexp} is the expansion but not the closed form of $q(x)$. This expansion is always obtainable from the expansion of $p(x)$ using the Lagrange inversion theorem. The second is that the expansions \eqref{eq:yexp} and \eqref{eq:zexp} are not necessarily always integer power series. For example, for metric functions which contain non-integer powers of $r$, this series might also contain non-integer powers,  as we will see for the Kiselev spacetime in Sec. \ref{ssec:apptoqs}. This, however, usually will not affect the integrability of the expanded series.

\section{$\Delta\phi$ and $\Delta t$ in the Kiselev BH spacetime\label{ssec:apptoqs}}

For the Kiselev BH spacetime, the metric functions in line element \eqref{eq:sssmetric} are \cite{Kiselev:2002dx}
\begin{equation}
A(r)=\frac{1}{B(r)}= 1-\frac{2M}{r}-\frac{\alpha}{r^{3{\omega}+1}} ,~C(r)=r^2. \label{eq:qmetric}
\end{equation}
Here $M$ is the spacetime mass, $\alpha$ is the parameter that controls the amount of the matter and $\omega<0$ is its EOS parameter in $P=\omega \rho$. In order to mimic the accelerated expansion of the universe, $\omega$ is chosen to be negative. Consequently, we have $\alpha>0$ in order for the matter energy density to be positive \cite{Kiselev:2002dx}. 
If $-1/3<\omega<0$, the spacetime is asymptotically flat and if  $-1\leq\omega< -1/3$ then the spacetime is non-asymptotically flat. In Subsec. \ref{subsec:aflat}, we will concentrate on the trajectory deflection and the total travel time of the former case while in Subsec. \ref{subsec:anonflat} the non-asymptotically flat case will be considered.

\subsection{Case  $-1/3< \omega< 0$\label{subsec:aflat}}

Using metric \eqref{eq:qmetric}, we can go through the procedure from Eq. \eqref{eq:pfuncdef} to \eqref{eq:dphifinal}. 
In particular, the function $p(x)$ in Eq. \eqref{eq:thetasdinp} is given by 
\be 
p(x)=v x \sqrt{\frac{1 -2 M x-\alpha  x^{1+3 \omega }}{v^2+\left(1-v^2\right) x \left(2 M+\alpha  x^{3 \omega }\right)}}.\label{eq:pxspec}\ee
And the expansion \eqref{eq:yexp} in this case is found to be
\be
y\lb \frac{u}{b}\rb =\sum_{m=0}^\infty \sum_{n=m}^\infty y_{m,n} \lb \frac{u}{b}\rb^{n+3m\omega }
\ee
with the first few of the coefficients
\begin{subequations}
\label{eq:ynmfirstfew}
\begin{align}
y_{0,0}=&1, \\
y_{0,1}=&M \lb 1+\frac{1}{v^2}\rb, \\ y_{0,2}=&\frac{3M^2}{2} \lb 1+\frac{4}{v^2}\rb , \\
y_{0,3}=&\frac{M^3}{2} \lb \frac{5}{2}+\frac{45}{2v^2}+\frac{15}{2v^4} -\frac{1}{v^6}\rb ,  \\
y_{1,1}=&\frac{\alpha}{2}\lb 1 +\frac{1+3\omega}{v^2} \rb, \label{eq:leadingalphay11} \\
y_{1,2}=&\frac{3\alpha M}{2} \lb 1+\frac{ \lb 4+6 \omega \rb}{v^2}+\frac{\omega \lb 2+3\omega \rb}{v^4}\rb, \\
y_{1,3}=&\frac{3\alpha M^2}{4}\lsb 5+\frac{45\lb 1+\omega \rb}{v^2}+\frac{15\lb 3 \omega ^2+4 \omega +1\rb}{v^4}\right.\nonumber\\
&\left.+\frac{9 \omega ^3+9 \omega ^2-\omega -1}{v^6} \rsb
\end{align}
\end{subequations}
Here there are two summation indices corresponding to integer powers of $u$ and $u^{3\omega}$ respectively. 
On the other hand, the expansion \eqref{eq:zexp} in this case is found to be
\be
z\lb \frac{u}{b}\rb =\sum_{m=0}^\infty \sum_{n=m-1}^\infty z_{m,n} \lb \frac{u}{b}\rb^{n+3m\omega }
\ee
with the first few of the coefficients
\begin{subequations}
\label{eq:znmfirstfew}
\begin{align}
z_{0,-1}=&\frac{1}{v}, \\
z_{0,0}=&M \lb \frac{ 3 }{v}-\frac{1}{v^3} \rb, \\ 
z_{0,1}=&\frac{15 M^2}{2 v}, \\
z_{0,2}=&\frac{M^3}{2} \lb \frac{35}{v}+\frac{35}{v^3}-\frac{7}{v^5} +\frac{1}{v^7}\rb ,  \\
z_{1,0}=&\frac{\alpha}{2}\lb \frac{3}{v} +\frac{3\omega-1}{v^3} \rb, \\
z_{1,1}=&\frac{3\alpha M}{2} \lb \frac{5}{v}+\frac{ 10\omega}{v^3}+\frac{\omega \lb 3\omega-2 \rb}{v^5}\rb, \\
z_{1,2}=&\frac{3\alpha M^2}{4}\lsb \frac{35}{v}+\frac{35\lb 1+3\omega \rb}{v^3}+\frac{7\lb 9 \omega ^2-1\rb}{v^5}\right.\nonumber\\
&\left.+\frac{9 \omega ^3-9 \omega ^2-\omega +1}{v^7} \rsb
\end{align}
\end{subequations}

Using Eqs. \eqref{eq:ynmfirstfew} and \eqref{eq:znmfirstfew}, the change of the angular coordinate in Eq. \eqref{eq:dphifinal} and total travel time in Eq. \eqref{eq:dtcase1exp} become
\begin{align}
&\Delta\phi = \sum_{i=s,d}\sum_{m=0}^\infty \sum_{n=m}^\infty \frac{y_{m,n}}{b^{n+3m\omega}} I_{m,n}(\theta_i),\label{eq:dphires1}\\
& \Delta t = \sum_{i=s,d}\sum_{m=0}^\infty \sum_{n=m-1}^\infty \frac{z_{m,n}}{b^{n+3m\omega}} I_{m,n-1}(\theta_i),\label{eq:dtres1}\end{align}
with $I_{m,n}(\theta_i)$ defined by
\be I_{m,n}(\theta_i)=\int_{ \sin\theta_i}^1\frac{u^{n+3m\omega}}{\sqrt{1-u^2}}\dd u~~~(i=s,~d)\ee
and their results are given in Eq. \eqref{eq:inmexp}. Here $\theta_{s,d}$ are still given by Eq. \eqref{eq:thetasdinp} with $p(x)$ in Eq. \eqref{eq:pxspec}. 
We comment that result \eqref{eq:dphires1} is the exact and complete change of the angular coordinate along the trajectory. Because the very weak dependence of $\theta_{s,d}$ on $b$, $\Delta \phi $ in Eq. \eqref{eq:dphires1} actually is a quasi-power series of $M/b$. 

Note that result \eqref{eq:dphires1} works for zero and positive $\omega$'s too because they also correspond to asymptotic spacetimes. When $\omega=0$, the metric \eqref{eq:qmetric} reduces to that of the Schwarzschild spacetime with $(M+\alpha/2)$ playing the role of the spacetime mass. We have checked that in this case, Eq. \eqref{eq:dphires1} agrees with the Schwarzschild spacetime result, i.e. Eq. (33) of \cite{Huang:2020trl}. When $\omega=1/3$, then deflection in RN spacetime, i.e. Eq. (23) of \cite{Xu:2021rld}, is recovered with $-\alpha$ replacing the charge square $Q^2$. We notice that Ref. \cite{Belhaj:2020rdb} computed the deflection angle of lightrays for source and observer at infinite radius to the first order of $\omega\to 0$. Unfortunately, a comparison shows that their result does not have the correct Schwarzschild limit at the $b^{-2}$ order. 

Later on, what will be used in the GL equation is $\Delta\phi$ in its pure series form of $M/b$. For this purpose, in Eq.  \eqref{eq:imnsmallthetaexp} we expanded $I_{m,n}(\theta_i)$ for small $\theta_i$ and then in Eq. \eqref{eq:imnsmallbori} further expand it using the relation \eqref{eq:thetasdinp} in terms of small $b/r_i$. Substituting them into Eq. \eqref{eq:dphires1}, then the result of $\Delta \phi$ to the leading non-trivial order of $M/b,~b/r_i$ and $\alpha$ becomes
\begin{align}
\Delta\phi=&\sum_{i=s,d}\lcb \lb \frac{\pi}{2}-\frac{b}{r_i}\rb +\frac{M}{b}\lb 1+\frac{1}{v^2}\rb \right.\nn\\
&\left.+\frac{\sqrt{\pi}\Gamma\lb 2+\frac{3\omega}{2}\rb}{(2+3\omega)\Gamma\lb \frac{3(1+\omega)}{2}\rb}\lb 1+\frac{1+3\omega}{v^2}\rb \frac{\alpha}{b^{1+3\omega}}\rcb\nonumber\\
&+\mathcal{O}(\text{higher orders}).
\label{eq:dphicase1exp}
\end{align}
Similarly, when computing the time delay, we will be needing the $\Delta t$ in the series form of $M/b$ and $b/r_i$. Substituting the expansions of $I_{m,n}$ into Eq. \eqref{eq:dtres1}, $\Delta t$ to the first three orders, i.e., orders $r_i,~b,~M$ and $\alpha$, becomes 
\begin{align}
\Delta t=&\sum_{i=s,d}\lcb \frac{r_i}{v}-\frac{b^2}{2r_i v}+\frac{\lb b^2+2 r_i^2\rb \lb2 M+\alpha  r_i^{-3 \omega }\rb }{4 r_i^2 v^3} \right.\nn\\
&\left.-\frac{M \lb 3 v^2-1\rb \lb b^2+4 r_i^2 \ln\frac{b}{2 r_i}\rb}{4 r_i^2 v^3}+\frac{\alpha\lb -1+3v^2+3\omega\rb}{12v^3}\right.\nn\\
&\left. \times \lsb \frac{3 \sqrt{\pi } \Gamma \left(\frac{3 \omega }{2}\right)}{b^{3 \omega } \Gamma \left(\frac{3 \omega }{2}+\frac{1}{2}\right)}-\frac{3 b^2 \omega +r_i^2 (6 \omega +4)}{\omega  (3 \omega +2) r_i^{3 \omega +2}}\rsb\rcb \nn\\
&+\mathcal{O}(\text{higher orders}).
\label{eq:dtcase1exp}
\end{align}
When $\omega=0$, Eq. \eqref{eq:dtcase1exp} reduces to the result in Schwarzschild spacetime, i.e. Eq. (35) of Ref. \cite{Liu:2020mkf}, with $(M + \alpha/2)$ playing the role of the spacetime mass. 

\begin{figure}[htp!]
\includegraphics[width=\columnwidth]{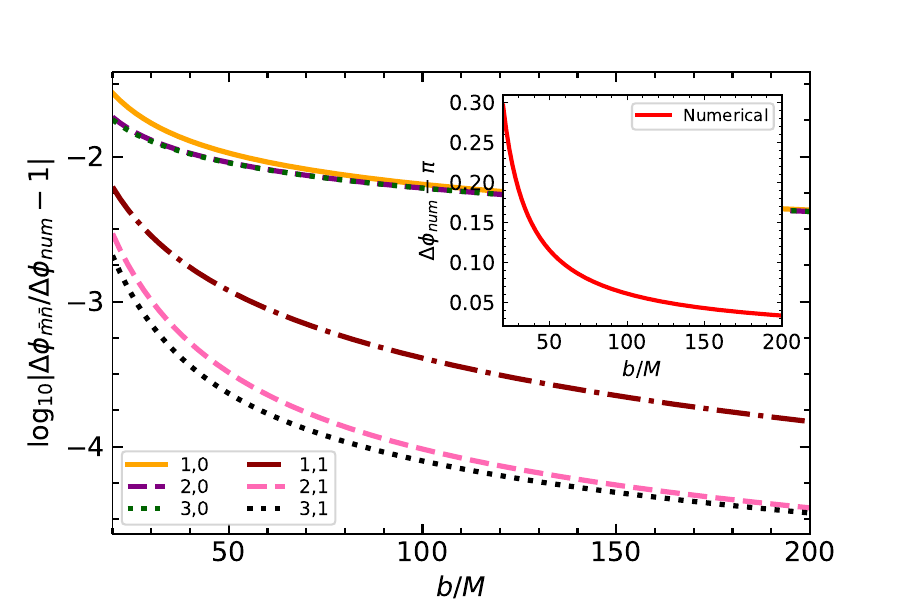}\\
(a)\\
\includegraphics[width=\columnwidth]{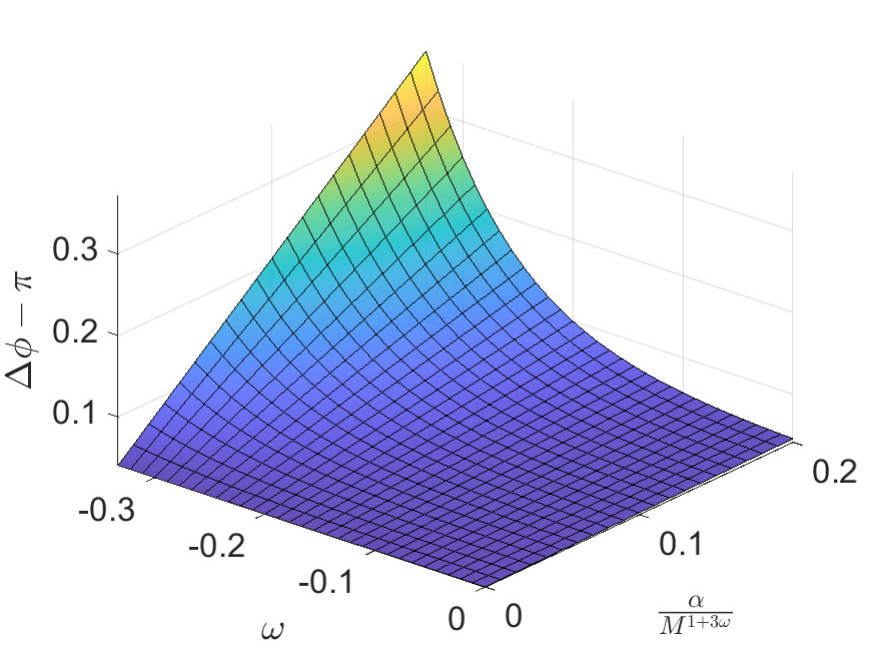}\\
(b)
\caption{\label{fig:dphiplot} 
(a) $\log|\Delta\phi_{\bar{m}\bar{n}}/\Delta\phi_{num}-1|$ as a function of $b$ from $20M$ to $200M$ for $\omega=-1/6,~\alpha=M^{3\omega+1}/10,~r_s=r_d=10^6M,~v=99/100$. Inset: the numerical integration result of Eq. \eqref{eq:deltaphi}. (b)
$\Delta\phi-\pi$ calculated using Eq. \eqref{eq:dphires1} as a function of $\alpha$ from 0 to $0.2M^{3\omega+1}$ and $\omega$ from -1/3 to 0 with $b=100M$. Other parameters are the same as in (a). 
}
\end{figure}

To verify the correctness of $\Delta\phi$ found in Eq. \eqref{eq:dphires1}, we can firstly define a truncated $\Delta\phi_{\bar{m}\bar{n}}$  
\begin{align} 
\Delta\phi_{\bar{m}\bar{n}} (b,\omega,\alpha)= \sum_{i=s,d} \sum_{m=0}^{\bar{m}}  \sum_{n=m}^{\bar{n}} \frac{y_{m,n}}{b^{n+3m\omega}} I_{m,n}(\theta_i),
\label{eq:dphiresbar}
\end{align}
and then see how they compare to deflection angle $\Delta\phi_\mathrm{num}$ obtained by direct numerical integration of the definition \eqref{eq:deltaphi}. As long as the numerical integration is done to high enough accuracy, then $\Delta\phi_\mathrm{num}$ can be thought of as the true deflection that $\Delta\phi_{\bar{m}\bar{n}}$ should approach. In Fig. \ref{fig:dphiplot} (a) we plot the difference of $\Delta\phi_{\bar{m}\bar{n}}$ and $\Delta\phi_\mathrm{num}$ as a function of $b$. It is seen that as the truncation order increases, the series result approaches the numerical integration very rapidly, and even more so for larger $b$. This is expected because the larger the $b$, the more accurate an inverse-$b$ series result such as Eq. \eqref{eq:dphires1} will be.  Also because of this, we expect that as $b$ increases, the $\Delta\phi$ should decrease monotonically. This is confirmed in the inset of Fig. \ref{fig:dphiplot} (a). Since $-1/3<\omega<0$ terms in Eq. \eqref{eq:dphiresbar} are proportional to $1/b^{n+3m\omega}$, it is understandable that the magnitude of each order will decrease smaller if $n$ increases by 1 than $m$ increases by 1. This is also reflected in Fig. \ref{fig:dphiplot} (a) that curves corresponding to the same $\bar{m}$ are closer to each other.

For the dependence of $\Delta\phi$ on other parameters, in this work we will concentrate on the effects of only $\alpha$ and $\omega$ since other parameters such as $v$ and $M$ have been well studied previously \cite{Jia:2020xbc}. For $\alpha$, we see from Fig. \ref{fig:dphiplot} (b) that $\Delta\phi$ for any fixed value of $\omega$ increases roughly linearly as $\alpha$ increases, just as revealed by the leading order contribution in Eq. \eqref{eq:dphicase1exp}. Also from the same term containing $\alpha$ in Eq. \eqref{eq:dphicase1exp}, we see that since $\alpha>0,~-1/3<\omega<0$ and $b\gg M$, the deflection will decrease as $\omega$ increases for any fixed $\alpha$.  This is actually a more unique effect  that few other works have studied because in most of those works the EOS parameter $\omega$ in the metric functions are fixed to particular values. 
Both the above effects indeed can be understood from the effective potential of the Kiselev BH spacetime. Using definition \eqref{eq:gveffdef}, this is
\be 
V_{\mathrm{eff}}(r)=\frac12\lb \frac{L^2}{r^2}+\kappa \rb\lb 1-\frac{2M}{r}-\frac{\alpha}{r^{1+3\omega}}\rb.
\label{eq:effinkb}
\ee
While the corresponding Eq. \eqref{eq:deltaphi} of $\Delta\phi$ in this case is given by
\be 
\Delta\phi=\sum_{i=s,d}\int_{r_0}^{r_i}\sqrt{\frac{L}{r\lsb E^2-2V_\mathrm{eff}(r)\rsb}} \dd r \label{eq:deltaphisubed}\ee
and the minimal approach $r_0$  is solvable from Eq. \eqref{eq:linr0}.
It is seen that for a fixed $b$ and $v$ which in turn fix $L$ according to Eq. \eqref{eq:leandbv}, the larger the $\alpha>0$ or the smaller the $\omega\in(-1/3,0)$, the smaller the potential \eqref{eq:effinkb} and the integrand of Eq. \eqref{eq:deltaphisubed}. 
However, the $r_0$ solved from Eq. \eqref{eq:linr0} for this case 
turns out to be smaller for larger $\alpha$ or smaller $\omega$, which then enlarges the integration range. It can be shown that the later factor actually wins the competition and therefore the entire $\Delta\phi$ becomes larger, as seen from Fig. \ref{fig:dphiplot} (b).

\subsection{Case $-1\leq \omega< -1/3$\label{subsec:anonflat}}

To fulfill the initially proposed purpose of such a spacetime, it is well known that $\omega$ should be close to $-1$ so that a cosmological constant term can be mimicked. Therefore from the application point of view, this range of $\omega$ is more important than the asymptotically flat one. 

Since in this case there exists a cosmological horizon, which can be recognized by inspecting the metric \eqref{eq:qmetric} for $\omega<-1/3$ and $\alpha>0$, the observer as well as the signals  will not be set to infinite $r$. Therefore, to compute the deflection angle \eqref{eq:deltaphi} in this case, one can not carry out the infinite $b$ (or $r_0$) expansion straightforwardly.  Besides, the physical meaning of $b$ as the distance from the asymptotics of the trajectory to its parallel radial direction is lost because of the asymptotic non-flatness, although one can still try to define an effective $b_{\mathrm{eff}}$ using \eqref{eq:leandbv}. 

We therefore have to use a different technique, which is developed in Ref. \cite{Li:2021qei}, to carry out a two-step expansion in the small  $\alpha$ limit first and then in the large $r_0$ limit.
By using this method, one can change the integration variable in Eq. \eqref{eq:deltaphi}  from $r$ to $u$
\begin{equation}
    	r\rightarrow u \cdot r_0,
\end{equation}
and then replace $r_0$ everywhere by 
\be
r_0\rightarrow \epsilon\alpha^{1/(1+3\omega)} \label{eq:r0toeps}
\ee 
where $\epsilon$ is a dimensionless and infinitesimal quantity.
We can expand the integrand of Eq. \eqref{eq:deltaphi} in small $\alpha$ first, and then in small $\epsilon$. Carrying out the expansions in this order allows $r_0$ to be large but not exceed the cosmological horizon. The result of the expansion of Eq. \eqref{eq:deltaphi}  is found to be
\begin{align}
 \Delta \phi =&  \sum_{i=s,d}\sum_{n=0}^{\infty} \frac{\alpha^{-n/(1+3\omega)}}{\epsilon^n} \sum_{m=0}^{\infty} \epsilon^{-m(1+3\omega)} \sum_{k=0}^{m+n} \int_{1}^{u_i}\dd u \nonumber \\
  & \times  \frac{G_{n,m,k}(u)E^{2k} \lb u+1 \rb^{m-k} \kappa^{1-\delta_{m+n,k} }  }{\lb E^2-\kappa\rb^{n+m} u^{n+1}(u^2-1)^{m+1/2}} , \label{eq:iniint}
\end{align}
where $u_i=r_i/r_0 ~(i=s,d)$ and the first several $G_{n,m,k}(u)$ are
\begin{subequations}\label{eq:gfirstfew}
\begin{align}
G_{0,0,0}(u) = & 1 ,\\
G_{0,1,0}(u) = & - \frac{\lb u-1 \rb u^{-1-3\omega}}{2} , \\
G_{0,1,1}(u) = & \frac{ u^2-u^{-1-3\omega} }{2} , \\
G_{1,0,0}(u) = & - M ,\\
G_{1,0,1}(u) = & M \lb u^2+u+1 \rb ,\\
G_{1,1,0}(u) = & \frac{ 3M \lb u-1 \rb u^{-1-3\omega} }{2} ,\\
G_{1,1,1}(u) = & - \frac{ M \lsb u^2 \lb 4u+1 \rb + u^{-1-3\omega} \lb u^3-6 \rb \rsb}{2} ,\\
G_{1,1,2}(u) = & \frac{ 3M \lb u^2+u+1 \rb \lb u^2-u^{-1-3\omega} \rb }{2} .
\end{align}
\end{subequations}
Higher order $G_{n,m,k}$'s can also be obtained without any difficulty. Moreover, one can show that since the integrands in Eq. \eqref{eq:iniint} are rational functions of $u$ and $u^{3\omega}$, the integration can always be carried out. Denoting the integration results as $I_{n,m,k}(u_i)~(i=s,~d)$ and substituting $\epsilon$ back to $\alpha$ and $r_0$ using Eq. \eqref{eq:r0toeps}, $\Delta\phi$ finally is computed to be
\begin{align}
\Delta \phi =&  \sum_{i=s,d}\sum_{n=0}^{\infty} \frac{1}{r_0^n} \sum_{m=0}^{\infty} \lb \frac{\alpha}{r_0^{1+3\omega}}\rb^m \sum_{k=0}^{m+n} I_{n,m,k}\lb \frac{r_i}{r_0}\rb \label{eq:dphirescase2}
\end{align}
where the first few $I_{n,m,k}$ corresponding to Eq. \eqref{eq:gfirstfew} are shown in Eq. \eqref{eq:ifirstfewmaple}.
In this result, clearly a nonzero $\alpha$ contributes to $\Delta\phi$ through the terms with $m\geq 1$ while the $m=0$ terms are the pure Schwarzschild contribution. 

Since in the current form of result \eqref{eq:dphirescase2}, the dependence of $\Delta\phi$ on the finite distance $r_s$ and $r_d$ are obscured by the hypergeometric functions in Eq. \eqref{eq:ifirstfewmaple}, it is also desirable to study the large $r_s$ and $r_d$ limit of Eq. \eqref{eq:dphirescase2}. Using the expansion of $I_{n,m,k}$ given in Eq. \eqref{eq:ifirstfewmapleexpand}, $\Delta\phi$ to the order of $(r_0/r_i)^1,~(M/r_0)^1$ and $\alpha^1$ becomes
\begin{align}
& \Delta\phi = \sum_{i=s,d} \lcb \frac{\pi }{2} - \frac{r_0}{r_i} + \frac{ \lb 2 E^2-\kappa \rb M}{ \lb E^2-\kappa \rb r_0}-\frac{E^2 M}{ \lb E^2-\kappa \rb r_i} \right. \nonumber\\
& + \lcb \frac{\kappa}{2\lb E^2-\kappa \rb \lb 2+3\omega \rb} \lsb  \lb\frac{ r_0}{r_i}\rb^{2+3\omega}  -\frac{\sqrt{\pi } \Gamma \lb 2+\frac{3\omega }{2}\rb  }{ \Gamma \lb \frac{3}{2}+\frac{3\omega }{2} \rb } \rsb \right. \nonumber \\
& \left.\left. +\frac{E^2}{E^2-\kappa}  \frac{\sqrt{\pi } \Gamma \lb 3+\frac{3\omega}{2}\rb  }{\lb 4+3\omega  \rb \Gamma \lb \frac{3}{2}+\frac{3\omega }{2}\rb }  \rcb \frac{\alpha}{r_0^{1+3\omega}} \rcb  + \mathcal{O} \lb \varepsilon^2 \rb \label{eq:c2daser}
\end{align}
where $\epsilon \sim \frac{M}{r_0},~\frac{r_0}{r_i},~\alpha$ and $\mathcal{O} \lb \varepsilon^2 \rb $ stands for the combined second-order infinitesimal.
Note the divergences of this as $\omega\to-2/3$ in the two factors of the first term are just artifacts and these two divergences actually cancel. 
We can compare this result with some works that dealt with a particular choice of $\omega=-2/3$ for lightrays with infinite source and detector distances. If we set $\kappa\to0,~\omega=-2/3$ and $r_{s,d}$ to infinity, terms with a negative power of $r_i$ in the above equation all vanish and Eq. (63) of Ref. \cite{Fernando:2012ue}, Eq. (39) of Ref. \cite{Amore:2006xp}, Eq. (39) (after expansion for small $c$ and $M/r_0$) of \cite{Malakolkalami:2015tsa},  Eq. (60) of \cite{Shchigolev:2016gro} and Eq. (48) (when $Q=0$) of Ref. \cite{Azreg-Ainou:2017obt},    
are recovered to the above order. Ref. \cite{Fernando:2012ue,Amore:2006xp} also agree with us at the order $\mathcal{O}(M^2/r_0^2),~\mathcal{O}(\alpha M)$ and $\mathcal{O}(\alpha^2 r_0^2)$. Eq. \eqref{eq:c2daser} for general $\omega$ but infinite $r_{s,d}$ and lightray also matches that of Ref. \cite{Azreg-Ainou:2017obt}. To be complete and for future reference for others, in Eq. \eqref{eq:defweyl} in Appendix \ref{app:b} we supplement the full result for $\omega=-2/3$ to the second combined order for null signals, with finite distance effect also taken into account. This is also exactly the deflection in Weyl gravity as we prove in Appendix \ref{app:b} that the $k$ parameter in Weyl gravity does not contribute to the total deflection angle when it is expressed in terms of $r_0$ \cite{Li:2021qei}. 

By using the same method as in $\Delta \phi$, one can change the integration variable in Eq. \eqref{eq:totaltime}  from $r$ to $u$, and then its new form becomes
\begin{align}
 \Delta t =&  \sum_{i=s,d}\sum_{n=0}^{\infty} \frac{\alpha^{-\frac{n-1}{1+3\omega}}}{\epsilon^{n-1}} \sum_{m=0}^{\infty} \epsilon^{-m(1+3\omega)} \sum_{k=0}^{m+n} \int_{1}^{u_i}\dd u \nonumber \\
  & \times  \frac{G^{'}_{n,m,k}(u)E^{2k+1} \lb u+1 \rb^{m-k} \kappa^{1-\delta_{m+n,k} }  }{\lb E^2-\kappa\rb^{n+m+1/2} u^{n-1}(u^2-1)^{m+1/2}} , \label{eq:iniinttt}
\end{align}
where the first several $G^{'}_{n,m,k}(u)$ are
\begin{subequations}\label{eq:gfirstfewtt}
\begin{align}
G^{'}_{0,0,0}(u) = & 1 ,\\
G^{'}_{0,1,0}(u) = & - \frac{3 \lb u-1 \rb u^{-1-3\omega}}{2} , \\
G^{'}_{0,1,1}(u) = & \frac{ 1+ \lb 2u^2-3 \rb u^{-1-3\omega} }{2} , \\
G^{'}_{1,0,0}(u) = & -3 M ,\\
G^{'}_{1,0,1}(u) = & M \lb 2u+3 \rb ,\\
G^{'}_{1,1,0}(u) = & \frac{ 15M \lb u-1 \rb u^{-1-3\omega} }{2} ,\\
G^{'}_{1,1,1}(u) = & - \frac{ M}{2} \lsb 4u+3 + \lb 20u^2+3u-30 \rb u^{-1-3\omega} \rsb ,\\
G^{'}_{1,1,2}(u) = & \frac{ M}{2} \lsb 4u^2+4u+3 \right. \nonumber\\ 
 & \left. + \lb 8u^3+8u^2-12u-15\rb u^{-1-3\omega} \rsb  .
\end{align}
\end{subequations}
Carrying out the integration with respect to $u$,
$\Delta t$ finally is found to be
\begin{align}
\Delta t =&  \sum_{i=s,d}\sum_{n=0}^{\infty} \frac{1}{r_0^{n-1}} \sum_{m=0}^{\infty} \lb \frac{\alpha}{r_0^{1+3\omega}}\rb^m \sum_{k=0}^{m+n} I^{'}_{n,m,k}\lb \frac{r_i}{r_0}\rb \label{eq:dtrescase2}
\end{align}
where the $I^{'}_{n,m,k}$ corresponding to Eq. \eqref{eq:gfirstfewtt} are shown in Eq. \eqref{eq:ifirstfewmaplett}. In large $r_s$ and $r_d$ limit, the total flight time to the order of $(r_0/r_i)^1,~(M/r_0)^1$ and $\alpha^1$ becomes
\begin{align}
 \Delta t = & \sum_{i=s,d} \frac{E}{\sqrt{E^2-\kappa}} \lcb r_i - \frac{r_0^2}{2r_i} + M \lb \frac{ 2E^2-3\kappa }{E^2-\kappa}\ln \frac{2r_i}{r_0} \right.\right. \nonumber \\
  & \left. \left. + \frac{E^2}{E^2-\kappa}  \rb - \frac{\alpha}{r_0^{3\omega}} 
  \lsb \frac{3\sqrt{\pi} \Gamma\lb 2+\frac{3\omega}{2} \rb}{\lb 2+3\omega \rb \Gamma\lb \frac{1}{2}+\frac{3\omega}{2} \rb} \right. \right. \nonumber \\
 & - \frac{\lb 2E^2-3\kappa\rb \sqrt{\pi} \Gamma\lb 1+\frac{\omega}{2} \rb}{3\lb E^2-\kappa\rb \omega \Gamma\lb -\frac{1}{2}+\frac{3\omega}{2} \rb} 
  \nonumber \\
 &  \left.\left. + \frac{2E^2-3\kappa}{6\lb E^2-\kappa\rb\omega} \lb \frac{r_i}{r_0}\rb^{-3\omega}\rsb\rcb + \mathcal{O} \lb \varepsilon^2 \rb . \label{eq:c2ttser}
\end{align}
When $\omega=-1$ and $\alpha=\Lambda/3$, this total travel time reduces to Eq. (39) of Ref. \cite{Li:2021qei}. We will use \eqref{eq:c2ttser} when computing the time delay in the case $-1\leq \omega<-1/3$.

\begin{figure}[htp!]
\includegraphics[width=\columnwidth]{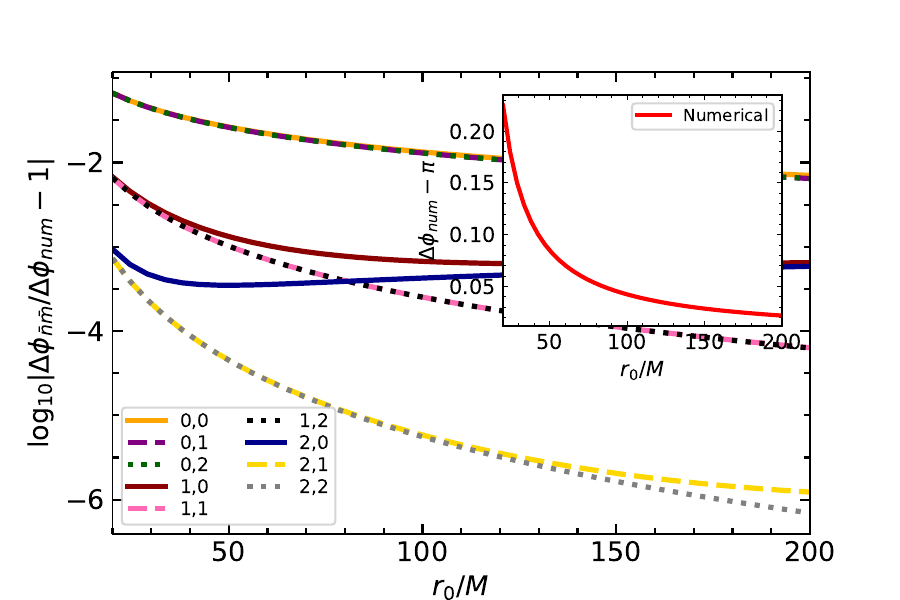}\\
(a)\\
\includegraphics[width=\columnwidth]{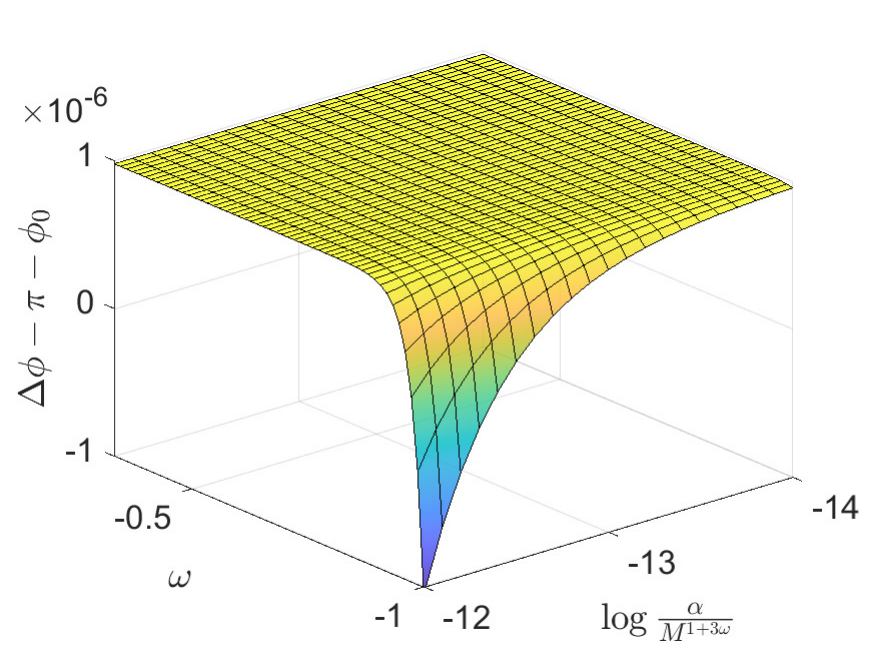}\\
(b)
\caption{\label{fig:dphiplotcase2} 
(a) $\log|\Delta\phi_{\bar{m}\bar{n}}/\Delta\phi_{num}-1|$ as a function of $r_0$ from $20M$ to $200M$ for $\omega=-1/2,~\alpha=10^{-13}M^{3\omega+1},~r_s=r_d=10^6M,~v=99/100$. The $\bar{n}=0$ lines almost overlap on the top, and so do the $\bar{n}=1,\bar{m}=0,1$ lines. Inset:the numerical integration result of Eq. \eqref{eq:deltaphi} for this case. (b)
$\Delta\phi-\pi-\phi_0$ calculated using Eq. \eqref{eq:dphirescase2} as a function of $\alpha$ from $10^{-14}M^{3\omega+1}$ to $10^{-12}M^{3\omega+1}$ and $\omega$ from $-1$ to $-1/3$ with $r_0=100M$ and $\phi_0=0.041$. Other parameters are the same as in (a). 
}
\end{figure}

To check the validity of $\Delta\phi$ given in Eq. \eqref{eq:dphirescase2} more thoroughly, similar to Eq. \eqref{eq:dphiresbar}, we can also construct a truncated $\Delta\phi_{\bar{n}\bar{m}}$ 
\begin{align}
&\Delta \phi_{\bar{n}\bar{m}} (r_0,\omega,\alpha)\nonumber\\
=&\sum_{i=s,d}\sum_{n=0}^{\bar{n}} \frac{1}{r_0^n} \sum_{m=0}^{\bar{m}} \lb \frac{\alpha}{r_0^{1+3\omega}}\rb^m  \sum_{k=0}^{m+n} I_{n,m,k}\lb \frac{r_i}{r_0}\rb \label{eq:dphirescase2trunc}
\end{align}
and then study its behavior against numerical integration results for some typical $-1\leq \omega<-1/3$. In Fig. \ref{fig:dphiplotcase2} (a) we plot 
$|\Delta \phi_{\bar{n}\bar{m}}/\Delta\phi_{\mathrm{num}}-1|$ of $r_0$. It is seen that similar to the case \ref{subsec:aflat}, as the truncation order increases, their difference diminishes very rapidly and even more so for larger $r_0$. Besides, from the inset figure we see the total deflection also decreases as $r_0$ increases, also as expected. Fig. \ref{fig:dphiplotcase2} (b) shows the dependence of $\Delta\phi$ on $\omega\in [-1,-1/3)$ and $\alpha$. We chose $\alpha$ to be a very small quantity because the cosmological horizon $r_H\approx \alpha^{1/(3\omega+1)}$ has to be larger than $r_s$ and $r_d$. It is seen that $\Delta\phi$ decreases monotonically as $\alpha$ increases or $\omega$ decreases. One can also understand these features from the effective potential \eqref{eq:effinkb}, $\Delta\phi$ in Eq. \eqref{eq:deltaphisubed} and the relation \eqref{eq:linr0}. 
When $\omega\in[-1,-1/3)$, the larger the $\alpha$ and the smaller the $\omega$, the smaller the third factor of the potential. And according to Eq. \eqref{eq:linr0} and using the metric \eqref{eq:qmetric}, for a fixed $r_0$, a larger $\alpha$ and smaller $\omega$ will result in a larger $L$ in the second factor of the potential and we can verify that the combined effect is that the entire potential becomes larger. Then Eq. \eqref{eq:deltaphisubed} implies that the deflection 
eventually becomes larger too. 

\section{Gravitational Lensing and time delay in the Kiselev BH spacetime}

To study the effect of the matter characterized by $\alpha$ and $\omega$ in the Kiselev BH spacetime on the GL, in this section we establish the GL equation in this spacetime and solve the apparent angles of the images. 

\begin{figure}[htp!]
\centering
\includegraphics[width=\columnwidth]{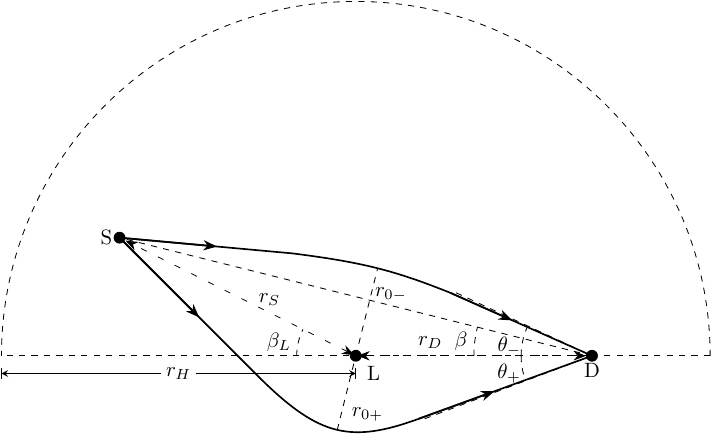}
\caption{The deflection and GL of signals. The counterclockwise and clockwise apparent angles are labeled as $\theta_+$ and $\theta_-$ respectively. The source (S) is located with an angle $\beta_L$ and $\beta$ with respect to the lens (L) and detector (D) against the lens-detector axis. The outer cosmological horizon with radius $r_H$ only exists for case 2.\label{fig:gl} }
\end{figure}

In this work, we will use an exact GL equation that was developed in Ref. \cite{Liu:2020wcu} and adopted in \cite{Li:2021qei}, which is particularly useful for deflection angles that take into account the finite distance effect. The exact GL equation involves the very definition of $\Delta\phi$ 
\be
\Delta\phi=\pi\pm\beta_L \label{eq:glgeneral}
\ee
where $\beta_L$ is the angle between the source and the lens-detector axis (see Fig. \ref{fig:gl}) and the $\pm$ signs correspond to the trajectory moving counter-clockwise and clockwise respectively. Note that $\beta_L$ can be exchanged with $\beta$ using Eq. \eqref{eq:blbrel}. From this equation, substituting Eq. \eqref{eq:dphicase1exp} for $\Delta\phi$ for the case $-1/3< \omega<0$ or Eq. \eqref{eq:c2daser}  for the case $-1\leq \omega<-1/3$, we can solve two impact parameters $b_\pm$ in the former case and two closest distances $r_{0\pm}$ for the latter case, corresponding to two trajectories in each case. 
With these $b_\pm$ or $r_{0\pm}$ known, then using formula \eqref{eq:thetasdinp}, \eqref{eq:dtcase1exp} and \eqref{eq:c2ttser}, the apparent angles and time delay can be readily obtained. 
In the following, we will show in more detail how to compute these quantities in each case.

\subsection{Case $-1/3< \omega< 0$ \label{subsec:case1}}

To get two impact parameters $b_\pm$, substituting Eq. \eqref{eq:dphicase1exp} into Eq. \eqref{eq:glgeneral} and keeping only the leading terms, the lensing equation in this case becomes
\begin{align}
 & \mp \beta_L -\frac{b}{r_s}-\frac{b}{r_d}+\frac{2M}{b}\lb 1+\frac{1}{v^2} \rb  \nonumber \\
  & + \frac{\sqrt{\pi}\Gamma\lb 2+\frac{3\omega}{2}\rb}{(2+3\omega)\Gamma\lb \frac{3(1+\omega)}{2}\rb}\lb 1+\frac{1+3\omega}{v^2}\rb \frac{\alpha}{b^{1+3\omega}} = 0 . \label{eq:glcase1subed}
\end{align}

This equation however can not be solved
analytically to get $b$ for general $\omega$.
To proceed, we have two options. The first is to look for the perturbative solution of $b$ in small $\alpha$. That is, when $\alpha$ is small, by supposing that the solution takes the form 
\begin{equation}
b_{\pm}(\alpha)= c_{b\pm0}+c_{b\pm1}\alpha+\calco(\alpha^2), \label{eq:bsmalla}
\end{equation}
we can use the method of undetermined coefficients in Eq. \eqref{eq:glcase1subed} to solve the coefficients $c_{b\pm0}$ and $c_{b\pm1}$ as
\bea 
c_{b\pm 0}&=& \mp \frac{\beta_L r_d r_s }{2  \lb r_d+r_s\rb}+\frac{\sqrt{r_d r_s}}{{2 v \lb r_d+r_s\rb}} \nonumber \\
  && \times  \sqrt{8 M \lb v^2+1\rb \lb r_d+r_s \rb +\beta_L ^2 r_d r_s v^2} , \label{eq:cb0}
\\
c_{b\pm 1}&=&\frac{c_{b\pm 0}^{1-3 \omega }r_d r_s}{c_{b\pm 0}^2 v^2 \lb r_d + r_s \rb +2 M r_d r_s \lb v^2+1\rb} \nonumber \\
  && \times \frac{\sqrt{\pi}\Gamma \lb \frac{3 \omega }{2}+2\rb}{\lb3 \omega +2\rb \Gamma \lb \frac{3 \lb \omega +1\rb}{2}\rb}\lb 1+v^2+3\omega\rb. \label{eq:cb1}
\eea
It is seen from Eq. \eqref{eq:cb0} that $c_{b\pm0}$ is $\omega$ independent and easy to verify that $c_{b\pm0}\gg M$. Then from Eq. \eqref{eq:cb1} we can check that for $\omega\in(-1/3,0)$, $c_{b\pm 1}$ is also positive but monotonically decreasing as $\omega$ increases towards zero. 
Therefore in this range of $\omega$, the desired $b$ for the signal to reach the observer is increased due to a positive $\alpha$, and the increased amount is smaller for larger $\omega$. 
This is understandable from the dependence of $\Delta\phi$ on $\alpha$ and $\omega$ as revealed in Eq. \eqref{eq:dphicase1exp} or the Fig. \ref{fig:dphiplot} (b). That is, for a fixed $\omega$ in this range, a positive $\alpha$ will cause a stronger deflection towards the lens, and the larger the $\omega$, the smaller the increase of the deflection. Therefore, for the signal to reach the same observer, its impact parameter has to be larger, but the amount of increase will be smaller for larger $\omega$. 

The second option for solving Eq. \eqref{eq:glcase1subed} is to try some specific $\omega$ for which Eq. \eqref{eq:glcase1subed} might be simplified to a polynomial equation of $b$. One such $\omega$ is $\omega=-1/6$ for which Eq. \eqref{eq:glcase1subed} becomes 
\bea
&&\mp \beta_L -\frac{b}{r_d}-\frac{b}{r_s}+\frac{2M}{b}\lb 1+\frac{1}{v^2}\rb\nn\\
&&+\frac{\sqrt{\pi}\Gamma\lb \frac{7}{4}\rb}{3\Gamma\lb \frac{5}{4}\rb}\lb 2+\frac{1}{v^2}\rb \frac{\alpha}{\sqrt{b}}=0
\eea
Its solution is found to be
\begin{align}
&b_{\pm}(\omega=-1/6)\nn\\
=&\frac14\lb  \sqrt{\frac{r_d r_s}{r_d+r_s} \lb \frac{2 g_1}{\sqrt{g_2}} \mp 2 \beta_L\rb-g_2 } +\sqrt{g_2}\rb^2 \label{eq:bom16}
\end{align} 
where
\begin{subequations}
\begin{align}
g_1=& \frac{\sqrt{\pi } \alpha   \Gamma \left(\frac{7}{4}\right)}{3 \Gamma \left(\frac{5}{4}\right)}\left(2+\frac{1}{v^2}\right),\\
g_2=&\frac{-r_dr_s}{3\lb r_d+r_s\rb}\left[ -\frac{24\sqrt[3]{2}M\lb r_d+r_s\rb}{r_d r_s \sqrt[3]{g_3}}\lb 1+\frac{1}{v^2}\rb \right .\nonumber\\
&\left. +\frac{\beta_L^2}{ \sqrt[3]{g_3}}-\frac{ \mp4 \beta_L-2^{2/3} \sqrt[3]{g_3}}{2}\right],\\
g_3=&g_4+\sqrt{g_4^2-4 \left[\beta_L^2-\frac{24M\lb r_d+r_s \rb}{r_d r_s}\lb1+\frac{1}{v^2} \rb\right]^3},\\
g_4=&\frac{9\lb r_d+r_s \rb}{r_d r_s}\left[ 3r_3^2\mp 16\beta_L M\lb1+\frac{1}{v^2} \rb \right]\mp 2\beta_L^3.
\end{align}
\end{subequations}
With this solution, in principle one can verify the dependence of the $b_\pm$ on larger $\alpha$. 

Substituting the results of $b_\pm$ in Eqs. \eqref{eq:bsmalla} or \eqref{eq:bom16}, and Eq. \refer{eq:pxspec} into \refer{eq:thetasdinp}, one can obtain the apparent angles $\theta_\pm$ of the two GL images at the detector
\begin{equation}
    \theta_\pm=\arcsin\lb b_\pm\cdot \frac{v}{r_d} \sqrt{\frac{r_d^{1+3\omega}-2 M r_d^{3\omega}-\alpha  }{\left(1-v^2\right) \left(2 M r_d^{3\omega}+\alpha\right)+r_d^{1+3\omega} v^2}} \rb.
    \label{eq:appangcase1}
\end{equation}
For the small $\alpha$ case, Eq. \eqref{eq:bsmalla} is valid only to the $\mathcal{O}(\alpha^1)$ order and therefore the apparent angle is also only accurate to this order. In addition, since $r_d\gg b_\pm\gg M$, we can also expand the result to the leading order of $b_\pm/r_{s,d}$ and $M/b_\pm$ and find 
\bea
    \theta_{\pm}&=&\frac{c_{b\pm 0}}{r_{d}}-\frac{Mc_{b\pm 0}}{r_{d}^2v^2}+\alpha\lb \frac{c_{b\pm 1}}{r_d}-\frac{c_{b\pm 0}r_d^{-2-3\omega}}{2v^2}\rb\nonumber\\
    &&+ \mathcal{O}\lb \frac{M^2}{r_d^2},\frac{\alpha Mc_{b\pm 1}}{r_d^2}\rb. \label{eq:thetapmsmallacase1}
\eea
We observe that in this equation the first term of the $\mathcal{O}(\alpha^0)$
order and the first term of the $\mathcal{O}(\alpha^1)$ order are nothing but $b_\pm/r_d$ with $b_\pm$ from Eq. \eqref{eq:bsmalla}, and the rest terms are the corrections introduced by the large square root term in Eq. \eqref{eq:mupmtrans}. Therefore one can expect that the effect of the parameters (such as $\beta,~\alpha,~\omega$) on $\theta_\pm$ are mainly determined by their effects on $b_\pm$.

\begin{figure}[htp!]
\includegraphics[width=\columnwidth]{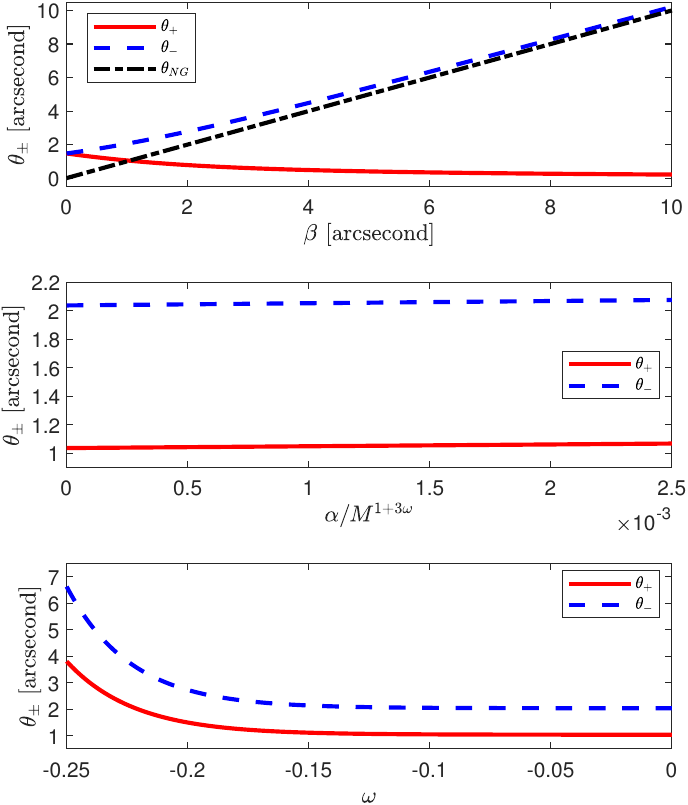}
\caption{\label{fig:thetamagcase1} 
The apparent angles $\theta_\pm$ given in Eq. \eqref{eq:thetapmsmallacase1}. Top: as a function of the source location $\beta$ from 0 to 10 [arcsecond] with $\omega=-0.1$ and $\alpha/M^{1+3\omega}=10^{-3}$; Middle: as a  function of $0<\alpha/M^{1+3\omega}\leq 2.5\times10^{-3}$ with $\omega=-0.1$ and $\beta=1$  [arcsecond] ; Bottom: as a function of $-0.25\leq \omega<0$ with $\beta=1$  [arcsecond] and $\alpha/M^{1+3\omega}=10^{-3}$. We used the Sgr A$^{*}$ SMBH data for $r_d=r_s=8.28$ [kpc] and $M=4.30\times10^6M_{\odot}$.
}
\end{figure}

\begin{figure}[htp!]
\includegraphics[width=\columnwidth]{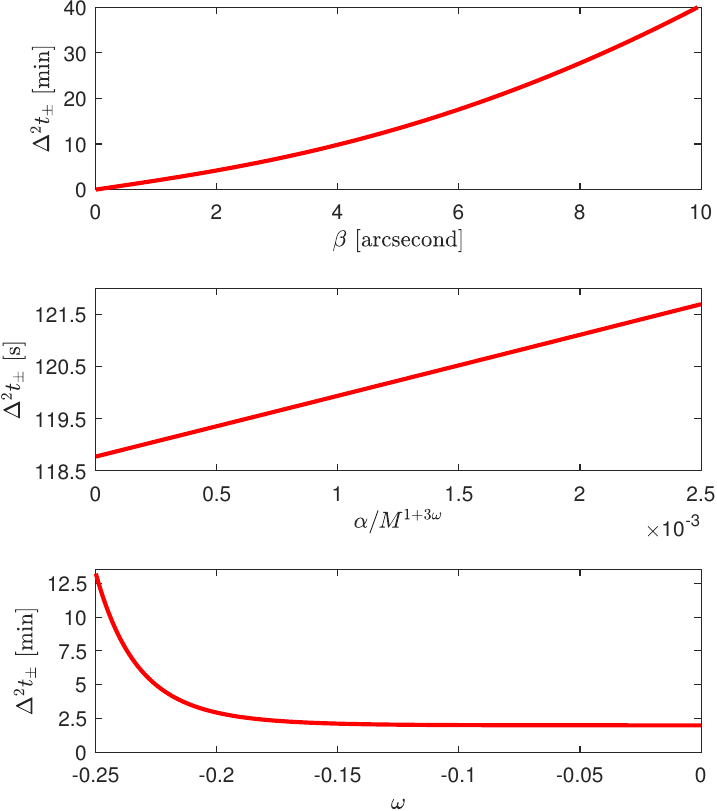}
\caption{\label{fig:timedelaycase1} 
The time delay $\Delta^2 t_\pm$ given in Eq. \eqref{eq:timedelaycase1}. Top: as a function of the source location $\beta$ from 0 to 10 [arcsecond] with $\omega=-0.1$ and $\alpha/M^{1+3\omega}=10^{-3}$; Middle: as a  function of $0<\alpha/M^{1+3\omega}\leq 2.5\times10^{-3}$ with $\omega=-0.1$ and $\beta=1$  [arcsecond] ; Bottom: as a function of $-0.25\leq \omega<0$ with $\beta=1$  [arcsecond] and $\alpha/M^{1+3\omega}=10^{-3}$. We used the Sgr A$^{*}$ SMBH data for $r_d=r_s=8.28$ [kpc] and $M=4.30\times10^6M_{\odot}$.
}
\end{figure}

To see these effects more clearly, in Fig. \ref{fig:thetamagcase1} we plot $\theta_\pm$
by assuming that the lens has a mass $M=4.30\times 10^6M_\odot$ of Sgr A$^{*}$ SMBH and $r_s=r_d=8.28$ [kpc] equals its distance to us \cite{GRAVITY:2021xju}. We assume that $\alpha=2.5\times10^{-3}M^{1+3\omega}$ which is within the uncertainty of the Sgr A$^{*}$ SMBH mass when $\omega=0$ \cite{GRAVITY:2021xju}. This mass ($4.30\pm 0.012\times 10^6 M_\odot$) is obtained by the {\it Gravity} team using the S star orbits. We use this value because it is much more accurate than the more recent value of $4.0^{+1.1}_{-0.6}\times 10^6 M_\odot$ obtained by fitting the Sgr A$^{*}$ SMBH shadow \cite{EventHorizonTelescope:2022xnr}. 
The uncertainty in this mass can be translated to the allowed value of $\alpha/M=0.012/4.3\approx 2.5\times 10^{-3}$ for the $\omega=0$ case of the Sgr A$^{*}$ SMBH, if it is a Kiselev spacetime. Therefore to respect this constraint, in all Figs. \ref{fig:thetamagcase1} to \ref{fig:timedelaycase2} in the following, we will restrict the range of $\alpha/M^{1+3\omega}$ to $2.5\times 10^{-3}$.

From the top panel of Fig. \ref{fig:thetamagcase1} we see that as $\beta$ deviates from zero, the $\theta_+$ and $\theta_-$ separate from each other with the former increasing while the latter decreasing. This is quite classical for all such GLs, as can be intuitively seen from the illustration Fig. \ref{fig:gl}: a larger $\beta$ would result in a larger $b_+$ and a smaller $b_-$ in order for the signal to still reach the detector. Therefore, $\theta_\pm$ have the above-mentioned changes as $\beta$ increases, which also agree qualitatively with other asymptotically flat spacetimes \cite{Liu:2015zou,Pang:2018jpm,Xu:2021rld}. The black dash-dot curve represents the apparent angle $\theta_{\mathrm{NG}}$ of the source when there is no gravity at all. Clearly, by the definition of $\beta$, this $\theta_{\mathrm{NG}}$ is nothing but $\beta$ itself, i.e., $\theta_{\mathrm{NG}}=\beta$. The fact that $\theta_-$ is always above  $\theta_{\mathrm{NG}}$ in this panel implies that the gravity in this case is always attractive so that the signal is always bent towards the lens.

For the effect of parameter $\alpha$ on $\theta_\pm$, from the middle panel we see that increasing $\alpha$ from zero will increase both apparent angles $\theta_+$ and $\theta_-$. It is easiest to understand this in the limit $\omega\to 0$, i.e., the Schwarzschild spacetime limit.
Then the deflection angle $\Delta\phi$ should also increase as $\alpha$ increases because $M+\alpha/2$ now plays the role of the lens mass. Consequently, in order for the signal to reach the same detector, both the impact parameters $b_\pm$ have to be larger, which results in larger $\theta_\pm$. For $\omega=-0.1$ as chosen in this panel, the qualitative effect of $\alpha$ are the same. We also note that in the plotted range of $\alpha$ which is fixed by the uncertainty of the Sgr A$^{*}$ SMBH mass, the effect of changing $\alpha$ is quite weak compared to the change of $\beta$ or $\omega$. 

Lastly for the effect of  $\omega$, a simple inspection of Eq. \eqref{eq:thetapmsmallacase1} shows that only the $\mathcal{O}(a^1)$ term depends on $\omega$ and its term containing $c_{b\pm 1}$ is much larger than the term involving $c_{b\pm 0}$. Then as revealed under Eq. \eqref{eq:cb1}, $c_{b\pm 1}$ increases monotonically like $\sim c_{b\pm 0}^{1-3\omega}$
as $\omega$ decreases from zero to $-1/3$. This explains the increase of $\theta_\pm$ for smaller $\omega$ in the bottom panel of Fig. \ref{fig:thetamagcase1}. 

With the two impact parameters of the two images known in Eq. \ref{eq:bsmalla}, substituting them into the total travel time \eqref{eq:dtcase1exp} and subtracting each other, one can obtain the time delay between these images
\begin{align}
\Delta^2 t_{\pm}&=\Delta t\lb b_{+}\rb -\Delta t\lb b_{-}\rb \nonumber\\
&= \sum_{i=s,d} \lcb \frac{M\lb 3v^2-1 \rb}{v^3} \ln \frac{b_-}{b_+}+\frac{3 \alpha  \lb v^2-1\rb \lb b_-^2-b_+^2\rb}{4 v^3 (3 \omega +2) r_i^{3 \omega +2}}\right. \nn \\ 
&\left.+\frac{\left(b_-^2-b_+^2\right) \lsb 3 M \left(v^2-1\right)+2 r_i v^2\rsb}{4 r_i^2 v^3}\right.\nn \\
&\left.+\frac{\sqrt{\pi } \alpha  \left(3 v^2+3 \omega -1\right) \Gamma \left(\frac{3 \omega }{2}\right) \left(b_+^{-3 \omega }-b_-^{-3 \omega }\right)}{4 v^3 \Gamma \left(\frac{3 \omega }{2}+\frac{1}{2}\right)}\rcb \nn \\
&+\mathcal{O}(\text{higher orders}).
\label{eq:timedelaycase1}
\end{align}
When $\omega=1/3$, Eq. \eqref{eq:timedelaycase1} agrees with time delay for neutral
particles in RN spacetime, i.e. Eq. (32) of Ref. \cite{Xu:2021rld}. When $\omega=0$, Eq. \eqref{eq:timedelaycase1} reduces to the time delay in Schwarzschild spacetime (see Eq. (45) of Ref. \cite{Liu:2020mkf}).

Similar to Fig. \ref{fig:thetamagcase1},
we plot the time delay \eqref{eq:timedelaycase1} in Fig. \ref{fig:timedelaycase1} to study its dependence on $\beta,~\alpha$ and $\omega$ using the same Sgr A$^{*}$ SMBH as the source.
From the top plot, it is seen that as $\beta$ increases or equivalently the two images become more separate from each other, the time delay monotonically increases to about 40 [min].
For the effect of $\alpha$, we see from the middle plot that for the entire plotted range of $\alpha$ the time delay increases by $\sim 3$ [sec], which is very small compared to its absolute value of about 2 [min]. This is also a reflection of the weak effect of $\alpha$ on $\theta_\pm$ as seen from Fig. \ref{fig:thetamagcase1}. While from the bottom plot, we see that as $\omega$ decreases from 0 to about $-1/6$, the time delay remains almost constant. When $\omega$ keeps decreasing to $-1/4$, the time delay increased to about 13 [min]. Both these two features agree with the observation of $\omega$'s effect on the image apparent angles $\theta_\pm$ in the last plot of Fig. \ref{fig:thetamagcase1}.  

For completeness, we also worked out the magnification in the small $\alpha$ limit in this case. The magnification of the images is defined as
\begin{equation}
    \mu_{\pm}=\left|\frac{\theta_{\pm}}{\beta}\frac{\dd \theta_{\pm}}{\dd \beta}\right|.\label{eq:magdef}
\end{equation}
where $\beta$ is the angle of the source if there were no lensing (see Fig. \ref{fig:gl}). To connect $\beta$ with $\beta_L$, the geometrical relation can be used
\begin{equation}
    r_s\sin\beta_L=\lb r_d+r_s\cos\beta_L\rb\tan\beta.\label{eq:blbrel}
\end{equation}
Then Eq. \eqref{eq:magdef} becomes 
\be
\mu_{\pm}=\left|\frac{\theta_{\pm}}{\beta}\frac{\dd \theta_{\pm}}{\dd \beta_L}\frac{\dd \beta_L}{\dd \beta}\right|. \label{eq:mupmtrans}
\ee
For the small $\alpha$ case, substituting Eq. \eqref{eq:thetapmsmallacase1}, the magnification is found as
\be
\mu_\pm
= \mu_{\pm,0}+\mu_{\pm,1}\alpha +\mathcal{O}\lsb \alpha^2,~\alpha \lb \frac{M}{r_d}\rb^{1/2}\rsb\label{eq:musmallacase1}
\ee
where
\bea
&&\mu_{\pm,0}= \frac{c_{b\pm 0}}{2r_d}\left|\frac{r_d  }{2c_{b\pm 0}\pm\beta r_d} \mp \frac{1}{\beta}\right|,\\
&&\mu_{\pm,1}= \frac{-c_{b\pm 1} v^2}{2r_d}\left| \frac{r_d  }{2c_{b\pm 0}\pm\beta r_d} \mp \frac{1}{\beta}\right|  \nonumber\\
&&\times  \lb \frac{3\omega-2}{v^2}  +\frac{2c_{b\pm 0}^2\lb r_d+r_s\rb }{c_{b\pm 0}^2\lb r_d+r_s \rb v^2+2Mr_dr_s\lb 1+v^2\rb}\rb.
\eea

\subsection{Case $-1\leq \omega<-1/3$}

Substituting Eq. \refer{eq:c2daser} into Eq. \refer{eq:glgeneral} and keeping only the leading terms, the lensing equation becomes
\begin{align}
& \mp \beta_L - r_0\lb \frac{1}{r_s}+\frac{1}{r_d} \rb + \frac{2M\lb 2E^2-\kappa \rb }{E^2-\kappa} \frac{1}{r_0} \nonumber \\
 & - \lb d_{1} r_0^{-1-3\omega} + d_{2} r_0 \rb\alpha  = 0, \label{eq:lenseqcase2}
\end{align} 
where
\begin{align}
    d_{1} = & \frac{\sqrt{\pi} }{\lb E^2-\kappa \rb \Gamma \lb \frac{3}{2}+\frac{3\omega}{2}\rb} \nonumber \\
    & \times \lsb \frac{2 E^2 \Gamma \lb 3 + \frac{3\omega}{2} \rb}{4+3\omega} - \frac{\kappa \Gamma \lb 2+\frac{3\omega}{2} \rb}{ 2+3\omega } \rsb \\
    d_{2} = & \frac{\kappa ( r_s^{-2-3\omega}+r_d^{-2-3\omega} ) }{2 \lb E^2-\kappa\rb \lb 2+3\omega \rb}.
\end{align}

Similar to the case $-1/3< \omega<0$, this equation can not be solved analytically for general $\omega $ either. However, there are still a few options we can consider. The first is again when $\alpha$ is small, we can solve $r_0$ as a series 
\begin{align}
    r_{0\pm}\lb \alpha\rb = c_{r_0\pm 0} + c_{r_0\pm 1} \alpha + \mathcal{O}\lb \alpha^2\rb, \label{eq:r0pmsol1}
\end{align}
where coefficients $c_{r_0\pm 0}$ and $c_{r_0\pm 1}$ can be determined by Eq. \refer{eq:lenseqcase2} as
\begin{align}
    c_{r_0\pm 0} = & \mp \frac{r_s r_d \beta_L}{2\lb r_s+r_d \rb} + \frac{\sqrt{r_s r_d}}{2\lb r_s+r_d \rb} \nonumber \\
    & \times \sqrt{ \beta_L^2 r_s r_d+ \frac{8M \lb r_s+r_d \rb \lb 2E^2 - \kappa\rb }{ E^2-\kappa} },
    \label{eq:c0def}\\
    c_{r_0\pm 1} = & -  \frac{\lb d_{1} c_{r_0\pm0}^{-2-3\omega} + d_{2} \rb r_s r_d c_{r_0\pm 0}^2 }{2\lb r_s+r_d \rb c_{r_0\pm 0} \pm \beta_L r_s r_d } .\label{eq:c1def}
\end{align}
Note from Eq. \eqref{eq:c0def} that since $r_{s,d}\gg M$, we have $r_{s,d}\gg c_{r_0\pm 0}\gg M$ regardless the value of $\beta_L$. 
The dependence of $r_{0\pm}$ on $\omega$ only appears in the $c_{r_0\pm 1}$ term but not the $c_{r_0\pm 0}$ term. In the null limit, $\kappa=0$ and the $d_2$ term drops out from Eq. \eqref{eq:r0pmsol1} and $d_1$ becomes
\begin{align}
    d_{1} = & \frac{4\sqrt{\pi}\Gamma \lb 3 + \frac{3\omega}{2} \rb }{(4+3\omega) \Gamma \lb \frac{3}{2}+\frac{3\omega}{2}\rb }. \label{eq:d1new}
\end{align}
If the SdS spacetime limit ($\omega\to -1$) is further taken, then since  $\Gamma[3/2+3\omega/2]\to \infty$, the $d_1$ term will disappear from $r_{0\pm}$ too. Then in this case the entire $r_{0\pm}$, as well as the original Eq. \eqref{eq:lenseqcase2}, will not depend on $\alpha$. This actually is in accord with the observation made in Ref. \cite{Li:2021qei} that for null signals in the SdS spacetime, its deflection angle will not depend on $\Lambda(=3\alpha)$ when expressed using $r_0$.

The second choice to solve Eq. \eqref{eq:lenseqcase2} is when $\omega$ is close to $-1$ and $\kappa=1$. In this case the $d_{2}$ term in Eq. \eqref{eq:lenseqcase2} is much larger than the $d_{1}$ term because $r_i\gg r_0$ and $\Gamma(0^+)\to\infty$. Keeping only the $d_{2}$ term, Eq. \eqref{eq:lenseqcase2} can be solved to find
\begin{align}
    & r_{0\pm}(\omega\to -1 ) = \frac{1}{2 \lb r_s+r_d-d_{2}r_sr_d \alpha \rb } \Bigg[ \mp r_s r_d \beta_L + \sqrt{r_s r_d} \nonumber \\
    & \left. \times \sqrt{ \beta_L^2 r_s r_d+ \frac{8M \lb r_s+r_d-d_{2}r_sr_d \alpha \rb \lb 2E^2 - \kappa\rb }{ E^2-\kappa} } \rsb. \label{eq:r0pmsol2}
\end{align}
Comparing this solution with Eq. (63) of Ref. \cite{Li:2021qei}, which considered the SdS spacetime case, we find that these two results will coincide exactly if we set $\omega=-1$ and $\alpha=\Lambda/3$ in Eq. \eqref{eq:r0pmsol2}.

There exists one more particular $\omega$ that allows the exact solution of Eq. \eqref{eq:lenseqcase2}, e.g., $\omega=-1/2$. For this choice, the GL equation \refer{eq:lenseqcase2} becomes a quartic polynomial of $r_0$ and its solution is
\begin{align}
    & r_{0\pm}(\omega=-1/2)=\frac14 \lb \sqrt{h_2} - \frac{d_1 \alpha}{2h_1} \right.\nonumber \\
    &\left.+  \sqrt{\frac{3 d_{1}^2\alpha^2\mp 8 \beta_Lh_1}{4 h_1^2}+ \frac{\pm 4 \beta_L d_{1} h_1 \alpha- d_{1}^3\alpha^3}{4 h_1^3 \sqrt{h_2}}-h_2}  \rb^2 ,
\end{align}
where
\begin{align}
    h_1 = & \frac{r_s+r_d}{r_sr_d}+d_{2} \alpha , \\
    h_2 = & \frac{3 d_{1}^2\alpha^2\mp 8 \beta_Lh_1}{6 h_1^2} + \frac{\sqrt[3]{2} \lsb \beta_L^2-12h_1M\lb 1+\frac{1}{v^2} \rb \rsb }{3h_1 \sqrt[3]{h_3}} \nonumber \\
    & - \frac{ \sqrt[3]{h_3}}{3 } , \\
    h_3 = & h_4 + \sqrt{h_4^2-4\lsb \beta_L^2-24h_1M\lb 1+\frac{1}{v^2} \rb \rsb^3 }, \\
    h_4 = & \pm 2\beta_L^3+18M\lb 1+\frac{1}{v^2} \rb\lb 8s\beta_Lh_1 - 3d_{1}^2\alpha^2 \rb.
\end{align}

\begin{figure}[htp!]
\includegraphics[width=\columnwidth]{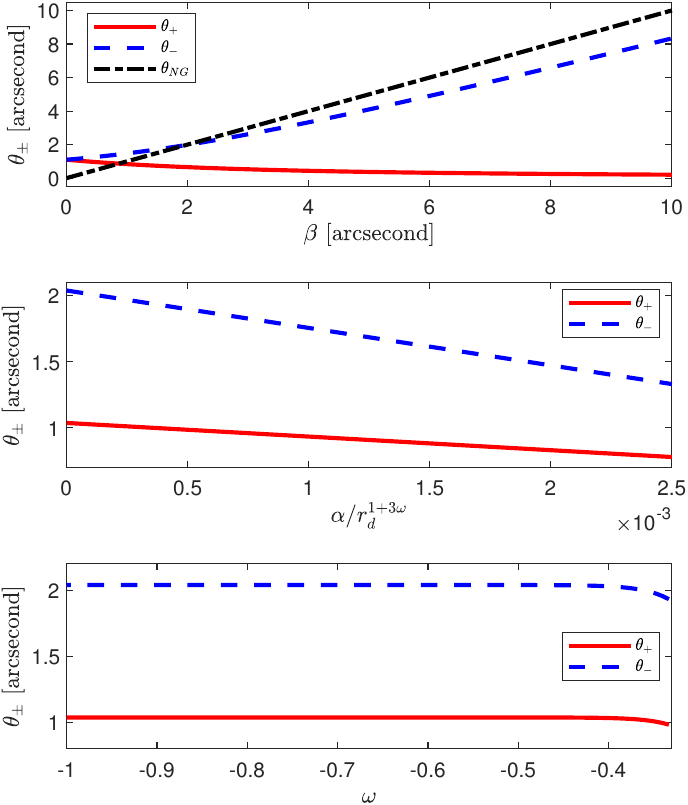}
\caption{\label{fig:thetamagcase2} 
The apparent angles $\theta_\pm$ given in Eq. \eqref{eq:thetapmsmalla}. Top: as a function of the source location $\beta$ from 0 to 10 [arcsecond] with $\omega=-1/2$ and $\alpha=2\times 10^{-3 }r_d^{1+3\omega}$; Middle: as a function of $\alpha\in [0, ~2.5\times 10^{-3 }r_d^{1+3\omega}]$ with $\beta=1$ [arcsecond] and $\omega=-1/2$; Bottom: as a function of $-1\leq \omega<-1/3$ with $\beta=1$ [arcsecond] and $\alpha=10^{-6 }r_d^{1+3\omega}$. 
In the bottom panel We choose the value of $\alpha$ to be a very small quantity because the cosmological horizon radius $r_H$ has to be larger than $r_{s,d}$ for the entire range of $\omega$. }
\end{figure}

With $r_{0\pm}$ known, substituting them and the metrics \eqref{eq:qmetric} into Eq. \eqref{eq:agform2}, we can obtain the apparent angles of the two images immediately 
\begin{align}
\theta_\pm =& \arcsin \lb \frac{r_{0\pm}}{r_d} \lcb \frac{\lb 2M-r_d+ \alpha r_d^{-3\omega} \rb}{ \lb 2M-r_{0\pm}+ \alpha r_{0\pm}^{-3\omega} \rb} \right. \right. \nonumber \\ 
  & \left.\left. \times \frac{\lsb 2M\kappa+(E^2-\kappa)r_{0\pm}+\alpha r_{0\pm}^{-3\omega}\rsb}{\lsb 2M\kappa+( E^2-\kappa)r_d + \alpha r_d^{-3\omega}\rsb } \rcb^{\frac{1}{2}} \rb \label{eq:thetapmcase2}
\end{align}
When $\alpha$ is small, then we should use Eq. \eqref{eq:r0pmsol1} for $r_{0\pm}$. Further noticing that $r_d\gg c_{r_0\pm0}\gg M$, apparent angle \eqref{eq:thetapmcase2} can be expanded to the leading orders of $\alpha,~c_{r_0\pm0}/r_{s,d}$ and $M/c_{r_0\pm0}$
to find
\begin{align}
\theta_{\pm} =& \frac{c_{r_0\pm 0}}{r_d} + \frac{E^2}{E^2-\kappa} \frac{M}{r_d}\nonumber \\
 & + \alpha \lsb \frac{c_{r_0\pm 1}}{r_d} - \frac{E^2 r_d^{-2-3\omega}  c_{r_0\pm 0}}{2 \lb E^2-\kappa \rb} + \frac{E^2 c_{r_0\pm 0}^{-3\omega} }{2r_d \lb E^2-\kappa \rb} \rsb  \nonumber\\
 &+ \mathcal{O}\lb \frac{c_{r_0\pm 0}M }{r_d^2},\frac{\alpha M^{-3\omega}}{r_d}, \alpha^2 \rb .
 \label{eq:thetapmsmalla}
\end{align}
When $\omega=-1$, Eq. \eqref{eq:thetapmsmalla} reduces to its corresponding results in the SdS spacetime, i.e. Eq.  (70) of Ref. \cite{Li:2021qei}.
One can verify after some simple algebra that the size of  the third term of $\alpha$'s coefficient is always much smaller than the second term except $\omega$ is extremely close to $-1/3$, at which point the last two terms cancel. Therefore, if $\omega$ is set to $-1/2$ as in the middle panel of Fig. \ref{fig:thetamagcase2},  the total coefficient of $\alpha$ will be negative (noticing $c_{r_0\pm 1}<0$) and both $\theta_+$ and $\theta_-$ will decrease as $\alpha$ increases. This dependence is easiest to understand by an analogy to the case $\omega=-1$  because it is known in the SdS spacetime that a positive $\Lambda$ will cause the decrease of the deflection angle \cite{Li:2021qei}. In other words, a small positive $\alpha=\Lambda/3$ effectively expels the signal a little so that the total deflection becomes smaller. Consequently, the apparent angles at both the source and the detector sides will have to be smaller in order for the signal to reach the same detector. 

In Fig. \ref{fig:thetamagcase2} we plot the apparent angles \eqref{eq:thetapmsmalla} to study its dependence on $\beta,~\alpha$ and $\omega$.  
From the top panel we see that qualitatively, the dependence of $\theta_\pm$ on $\beta$ has the same form as in the asymptotically flat $\omega\in(-1/3,~0]$ case (as least for the chosen ranges of parameters): the $\theta_-$ increases and $\theta_+$ decreases as $\beta$ increases. However, one sees that unlike the former case, as $\beta$ reaches a relatively large value of 10 [arcsecond], the larger apparent angle $\theta_-$ only reaches about 8.3 [arcsecond], quite far from $\beta$ itself. Again, this is because when $\beta$ is this large, the corresponding $b_-$ is also large and the top trajectory in Fig. \ref{fig:gl} has been bent away from the lens by the positive $\alpha$, and consequently the $\theta_-$ would be smaller than $\beta$. Indeed, we can read off the actual value $\beta_c$ at which the attraction due to $M$ and repulsion due to $\alpha$ cancel each other from the intersection of the no-gravity apparent angle $\theta_{\mathrm{NG}}$ (which exactly equals $\beta$) with the original $\theta_-$ curve. We see that this value is about $\beta_c\approx2.1$ [arcsecond]. In other words, if $\beta<\beta_c$ (or $\beta>\beta_c$), the attraction of $M$ will be stronger (or weaker) than the repulsion of $\alpha$ and the signal is bent towards (or away from) the lens.

For the effect of $\alpha$, we have argued under Eq. \eqref{eq:thetapmsmalla} that the coefficient of $\alpha$ is negative and consequently the larger the $\alpha$ the smaller the $\theta_\pm$, as shown in the middle panel. This also agrees qualitatively with the findings in SdS spacetime \cite{Li:2021qei}. Lastly the last panel shows the effect of $\omega$ which is uniquely studied in this paper. It is seen that as $\omega$ decreases from $-1/3$ to $-1$, both apparent angles $\theta_\pm$ increase to their asymptotic value. The reason for this behavior traces back to a detailed comparison of the three terms of $\alpha$'s coefficient in Eq. \eqref{eq:thetapmsmalla}. A simple numerical analysis shows that for the given choice of other parameters, the first term of $\alpha$'s coefficient is larger (or smaller) than the second term if $-2/3\leq\omega<-1/3$ (or $-1\leq\omega<-2/3$), while the third term is always dominated by either the second or the first terms. Then when $\omega$ is close to $-1$, the dependence of $\theta_\pm$ on $\omega$ will be  proportional to $\alpha r_d^{-2-3\omega}\propto r_d^{-1}$ since $\alpha=10^{-6}r_d^{1+3\omega}$ in this plot. That is, both $\theta_\pm$ should be flat when $\omega\to -1$. When $\omega$ approaches $-1/3$ from below, since the first term is larger than the second one (which is flat) in size, the total $\alpha$ correction to $\theta_\pm$ should be increasing in size with a negative coefficient. That is, both $\theta_\pm$ decreases. 

By using $r_{\pm}$ in \refer{eq:r0pmsol1} and total flight time \refer{eq:c2ttser}, the time delay between two images can be obtained as
\begin{align}
 \Delta^2 t_{\pm} =& \Delta t(r_{0+})-\Delta t(r_{0-}) \nonumber \\
 = & \sum_{i=s,d} \frac{E}{\sqrt{E^2-\kappa}} \lcb \frac{r_{0-}^2-r_{0+}^2}{2r_i} + \frac{ 2E^2-3\kappa }{E^2-\kappa} M \ln \frac{r_{0-}}{r_{0+}} \right. \nonumber \\
  & \left.  + \alpha \lb \frac{1}{r_{0-}^{3\omega}} - \frac{1}{r_{0+}^{3\omega}} \rb 
  \lsb \frac{3\sqrt{\pi} \Gamma\lb 2+\frac{3\omega}{2} \rb}{\lb 2+3\omega \rb \Gamma\lb \frac{1}{2}+\frac{3\omega}{2} \rb} \right. \right. \nonumber \\
 &  \left. \left. - \frac{\lb 2E^2-3\kappa\rb \sqrt{\pi} \Gamma\lb 1+\frac{\omega}{2} \rb}{3\lb E^2-\kappa\rb \omega \Gamma\lb -\frac{1}{2}+\frac{3\omega}{2} \rb} 
 \rsb \rcb + \mathcal{O} \lb \varepsilon^2 \rb. \label{eq:c2tdser}
\end{align}
After setting $\omega=-1$ and $\alpha=\Lambda/3$, this result agrees with the Schwarzschild-(a)de Sitter case found in Eq. (78) of Ref. \cite{Li:2021qei}.

\begin{figure}[htp!]
\includegraphics[width=\columnwidth]{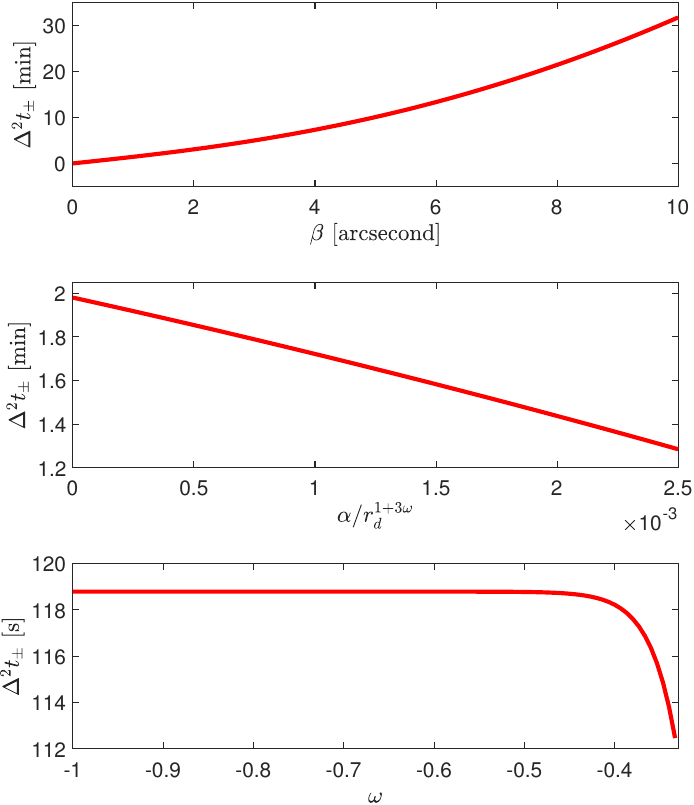}
\caption{\label{fig:timedelaycase2} 
The time delay $\Delta^2 t_{\pm}$ given in Eq. \eqref{eq:c2tdser}. Top: as a function of the source location $\beta$ from 0 to 10 [arcsecond] with $\omega=-1/2$ and $\alpha=2\times 10^{-3 }r_d^{1+3\omega}$; Middle: as a function of $\alpha\in [0, ~2.5\times 10^{-3 }r_d^{1+3\omega}]$ with $\beta=1$ [arcsecond] and $\omega=-1/2$; Bottom: as a function of $-1\leq \omega<-1/3$ with $\beta=1$ [arcsecond] and $\alpha=10^{-6 }r_d^{1+3\omega}$. 
In the bottom panel, We choose the value of $\alpha$ to be a very small quantity because the cosmological horizon radius $r_H$ has to be larger than $r_{s,d}$ for the entire range of $\omega$. }
\end{figure}

To reveal the effect of parameters $\beta,~\alpha$ and $\omega$ in this case, we plot in Fig. \ref{fig:timedelaycase2}
the time delay \eqref{eq:c2tdser} between images using Sgr A$^{*}$ SMBH as the lens. For the lens angle $\beta$, we see that qualitatively, its effect is similar to the case of $-1/3<\omega<0$. As $\beta$ increases, the time delay also increases monotonically, but only to about 30 [min]. This agrees with $\beta$'s effect on the image apparent angles $\theta_\pm$ and the fact that in this case, $\theta_+$ and $\theta_-$ are only separated by about $\sim$8 [arcsecond] when $\beta$ reaches its maximum value, comparing the $\sim$10 [arcsecond] separation in the $-1/3<\omega<0$ case.
For the effect of $\alpha$, we see from the middle plot that as $\alpha$ increases, the time delay decreases monotonically and almost linearly by a sensible fraction ($\sim$2 [min] to $\sim$1.3 [min]). This is also in accord with the effect of $\alpha$ on $\theta_\pm$ in Fig. \ref{fig:thetamagcase2}. Lastly, for the effect of $\omega$, as it decreases it is seen that it mainly causes a small increase of the time delay from 112 [second] to 118 [second] before reaching $-0.4$, after which the time delay remains almost a constant. Again, this can be understood from the effect of $\omega$ on the images' apparent angles $\theta_\pm$ as shown in the bottom plot of Fig. \ref{fig:thetamagcase2}. 

To be complete, we also computed the magnifications of the images in this case from Eq. \eqref{eq:mupmtrans} but with $\theta_\pm$ given by Eq. \eqref{eq:thetapmcase2}. Again, we  concentrate on the small $\alpha$ case, in which $\theta_\pm$ can be further replaced by Eq. \eqref{eq:thetapmsmalla}. Then the result is found to be 
\be
\mu_{\pm} = \mu_{\alpha,0} + \mu_{\alpha,2} \alpha + \mathcal{O} \lb \alpha^2 \rb, \label{eq:musmallacase2}
\ee
where
\begin{align}
\mu_{\alpha,0} =& \frac{(r_d+r_s)^2 c_{r_0\pm} c_{r_0\pm}^\prime }{\beta_L r_d^2 r_s^2 } ,  \\
\mu_{\alpha,1} =& \frac{\lb r_s+r_d \rb^2}{2\beta_L r_s^2 r_d^3} \lcb \frac{2 r_s r_d^2 c_{r_0\pm}^{1-3\omega} \lb d_{1}+d_{2} c_{r_0\pm}^{2+3\omega}\rb }{ \lsb \pm \beta_L r_s r_d +2 c_{r_0\pm}  ( r_s+r_d )\rsb^2} \right. \nonumber \\
 & \times \lsb \pm r_s r_d + 2 ( r_s+r_d ) c_{r_0\pm}^\prime \rsb  \nonumber \\
 &  - \frac{2 r_s r_d^2 c_{r_0\pm}^{-3\omega} c_{r_0\pm}^\prime \lsb 3 d_{2} c_{r_0\pm}^{3\omega  +2}+ \lb 1-3 \omega \rb d_{1} \rsb}{ \pm \beta_L r_s r_d +2 c_{r_0\pm}  ( r_s +r_d ) }  \nonumber \\
  & \left.  -  \frac{2 c_{r_0\pm}  E^2 r_d^{-3 \omega  }}{E^2-\kappa }  c_{r_0\pm}^\prime \rcb ,\\
c_{r_0\pm}^\prime = &\frac{\dd c_{r_0\pm}(\beta_L)}{\dd\beta_L}= \frac{\beta_L r_s r_d}{2 \lb r_s+r_d \rb} \lsb  \mp \frac{1}{\beta_L} \right. \nonumber \\
 & \left. + \lb \beta_L^2+ \frac{8M \lb r_s+r_d \rb \lb 2E^2-\kappa\rb}{r_s r_d \lb E^2-\kappa\rb } \rb^{-\frac{1}{2}} \rsb.
\end{align}

\section{Conclusions and discussion}

In this work, we used perturbative method to find the deflection angles of both null and timelike signals in the Kiselev BH spacetime with a variable EOS parameter $\omega$ with the finite distance effect of the source and detector taken into account. Although the fundamental principles of the perturbation method are similar, the technical details are different for the asymptotically flat case of $-1/3<\omega<0$ and the non-asymptotically flat case of $-1\leq\omega<-1/3$.  For the former case, the deflection angles obtained in Eq. \eqref{eq:dphires1} has a quasi-power series form of $b^{-(n+3m\omega)}$ and can be further series expanded in the small $b/r_{s,d}$ limit to get its expansion \eqref{eq:dphicase1exp}. For the latter case, the deflection angle in Eq. \eqref{eq:dphirescase2} takes a dual series form of $r_0^{-[n+(1+3\omega)m]}$ and $\alpha^m$ and its small $r_0/r_{s,d}$ expansion is given in Eq. \eqref{eq:c2daser}. It is found that for the former case (or the latter case), when $\alpha$ increases or $\omega$ decreases, the deflection angle will increase (or decrease). These results and features are verified numerically in Figs. \ref{fig:dphiplot} and \ref{fig:dphiplotcase2}.

Since these deflection angles naturally take into account the finite distance of the source and detector, an accurate lensing equation is used to solve the apparent angles $\theta_\pm$. The perturbative results of $\theta_\pm$ for small $\alpha$ for the two cases are given in Eqs. \eqref{eq:thetapmsmallacase1} and \eqref{eq:thetapmsmalla} respectively, and the effect of the parameters $\alpha$ and $\omega$ are analyzed. It is found that for the asymptotically flat case ($-1/3<\omega<0$), increasing $\alpha$ or decreasing $\omega$ would cause an increase of the apparent angles. While for non-asymptotically flat case ($-1\leq\omega<-1/3$), increasing $\alpha$ or $\omega$ will both lead to smaller apparent angles. These features can be understood qualitatively using their more familiar limits, i.e. the Schwarzschild limit with $\omega=0$ and SdS limit with $\omega=-1$.

If we consider the results in Fig. \ref{fig:thetamagcase1} to \ref{fig:timedelaycase2} from an observational point of view, 
observables in these figures offer a new and quantitative way to constrain the value of $\alpha$ and $\omega$ in the Kiselev spacetime, providing images of such sources can be observed in the future. In summary, from these plots we found that for Sgr A$^{*}$ SMBH and the case $-1/3<\omega<0$, the apparent angles $\theta_\pm$ can constrain the value of $\alpha$ very poorly while the time delay is more strongly affected by $\alpha$. For the case $-1\
leq\omega<-1/3$, both the apparent angles and the time delay are affected by $\alpha$ in a greater way compared to the smaller $|\omega|$ case. 

Finally, let us also comment on the possibility to use the EHT results on the M87$^{*}$ \cite{EventHorizonTelescope:2019dse,EventHorizonTelescope:2019ggy} and Sgr A$^{*}$ \cite{EventHorizonTelescope:2022xnr} SMBH shadows to constrain the values of $\alpha$ and $\omega$ of the Kiselev spacetime. Indeed as done in Ref. \cite{EventHorizonTelescope:2022xqj}, only after performing numerical simulations of the SMBH shadows for different values of $(\alpha,~\omega)$ and comparing to the ETH observed shadows directly, can one obtain information about $(\alpha,~\omega)$ from these shadows. Performing such simulations however, is beyond the scope of the current paper. The fundamental reason is that the physics in EHT black hole shadows happens mainly in the strong field regime of gravity, i.e., roughly around/between the innermost stable circular orbit (radius $\sim6M$) and the photon sphere (radius $\sim 3M$). While the methodology and results in this work mainly concerns the physics in the weak field limit of gravity ($r_0\sim b\gg M$), and are only applicable in this limit too.

\begin{acknowledgements}
We thank Mr. Ke Huang for his valuable discussions. This work is supported by the MOST China 2021YFA0718500.
\end{acknowledgements}

\begin{widetext}

\appendix
\section{Integration formulas and deflection for $\omega=-2/3$ \label{sec:appd}}

In the computation of $\Delta\phi$ in Sec. \ref{ssec:apptoqs}, the following integral formula is needed
\be
I_{m,n}(\theta_i)=\int_{ \sin\theta_i}^1\frac{u^{n^\prime}}{\sqrt{1-u^2}}\dd u=\frac{u^{n^\prime+1}}{n^\prime+1} \, _2F_1\left[\frac{1}{2},\frac{n^\prime+1}{2},\frac{n^\prime+3}{2},u^2\right]\Bigg|_{ \sin\theta_i}^1~~~~(m\geq0,~n\geq m).
\label{eq:inmexp} \ee
where $n^\prime=n+3m\omega$ and
the hypergeometric function $_2F_1$ is defined as 
\be
_2F_1[a,b,c,z] =\frac{\Gamma[c]}{\Gamma[b]\Gamma[c-b]}
\int_{0}^{1}t^{b-1}(1-t)^{c-b-1}(1-tz)^{-a}\dd t.
\ee
The first few $I_{m,n}$ can be slightly simplified using more elementary functions, to be
\begin{subequations}\label{eq:ifirstfewmaple2ind}
\begin{align}
I_{0,-2}(\theta_i) = & \cot\theta_i, \\
I_{0,-1}(\theta_i) = & \ln\lsb\cot{\frac{\theta_i}{2}}\rsb, \\
I_{0,0}(\theta_i) = & \frac{\pi}{2} - \theta_i, \\
I_{0,1}(\theta_i) = & \cos{\theta_i}, \\
I_{0,2}(\theta_i) = & \frac{\pi}{4}-\frac{\theta_i}{2}+\frac{\sin{\theta_i}\cos{\theta_i}}{2}, \\
I_{0,3}(\theta_i) = & \frac{\cos{\theta_i}}{3}\lb 2+\sin^2{\theta_i}\rb, \\
I_{1,n}(\theta_i)=&\frac{1}{n+1+3\omega}\lsb \frac{\sqrt{\pi } \Gamma \left(\frac{3 \omega+n+3 }{2}\right)}{\Gamma \left(\frac{3 \omega+n+2 }{2}\right)}-\sin ^{3 \omega +n+1}{\theta_i} \, _2F_1\left(\frac{1}{2},\frac{3
\omega+n+1 }{2};\frac{3 \omega+n+3 }{2};\sin ^2{\theta _i}\right)\rsb.
\end{align}
\end{subequations}
where $\Gamma(x)$ is the $\Gamma$ function. These formulas allow us to express the deflection angle to the first few orders in elementary functions and further expand in other small quantities such as $b/r_{s,d}$.

For small $\theta_i$, the result \eqref{eq:inmexp} then can be expanded as
\be
I_{m,n}(\theta_i)=\frac{\sqrt{\pi}\Gamma\lb \frac{n^\prime+3}{2}\rb }{(n^\prime+1)\Gamma\lb \frac{n^\prime+2}{2}\rb}+\sin^{n^\prime}\theta_i\lsb -\frac{\sin\theta_i}{n^\prime+1}-\frac{ \sin\theta_i^3}{2(n^\prime+3)}+\mathcal{O}\lb\sin^4\theta_i\rb\rsb.
\label{eq:imnsmallthetaexp}
\ee

We can further use the relation \eqref{eq:thetasdinp} to expand this to leading orders of $b/r_i$, which becomes
\be
I_{m,n}\lsb \theta_i ( \frac{b}{r_i}) \rsb=\frac{\sqrt{\pi}\Gamma\lb \frac{n^\prime+3}{2}\rb }{(n^\prime+1)\Gamma\lb \frac{n^\prime+2}{2}\rb}-\lb  \frac{1}{ 1+n^\prime}-\frac{M }{v^2r_i}-\frac{\alpha }{2v^2r_i^{1+3\omega}}+\frac{b^2}{2(3+n^\prime)r_i^2}-\frac{b^2M}{2v^2r_i^3}-\frac{\alpha b^2}{4v^2r_i^{3+3\omega}}\rb \lb \frac{b}{r_i}\rb^{1+n^\prime}+\mathcal{O}\lb \frac{1}{r_i^4}\rb.
\label{eq:imnsmallbori}
\ee
In particular, when $n^\prime=-1$, we can get the following limit of this expansion
\be
I_{m,n}\lsb \theta_i \lb \frac{b}{r_i}\rb \rsb=\ln\frac{2r_i}{b}+\frac{M}{r_i v^2}+\frac{\alpha}{2v^2 r_i^{1+3\omega}}-\lb\frac{1}{4}-\frac{M}{2r_iv^2}-\frac{\alpha}{4v^2r_i^{1+3\omega}} \rb \lb \frac{b}{r_i} \rb^2 +\mathcal{O}\lb \frac{1}{r_3^2}\rb
\label{eq:imnsmallbori_lim}
\ee

If $\omega=0$, then the integral \eqref{eq:inmexp} is equivalent to the $m=0$ case and then the integration can be carried out using a change of variables $u=\sin\xi$ to find an elementary expression
\be
I_{0,n}(\theta_i)=\int_{\sin \theta_i }^1\frac{u^n}{\sqrt{1-u^2}}\dd u=\int_{\theta_i}^{\pi/2}\sin^m\xi\dd \xi 
=
 \frac{(n-1)!!}{n!!}\times\begin{cases}
\displaystyle  \left(\frac{\pi}{2}-\theta_i
    +\cos\theta_i\sum_{j=1}^{[\frac{n}{2}]} \frac{(2j-2)!!} {(2j-1)!!}\sin^{2j-1} \theta_i\right),\\
~~~~~~~~~~~~~~~~~~~~~~~~~~~~n>0\text{ is even},\\
\displaystyle  \cos\theta_i \left(1
    +\sum_{j=1}^{[\frac{n}{2}]} \frac{(2j-1)!!}{(2j)!!} \sin^{2j}\theta_i\right),\\
~~~~~~~~~~~~~~~~~~~~~~~~~~~~n>0\text{ is odd}.
\end{cases} 
\label{eq:inthetares}
\ee
In particular, when $\theta_i=0$ as in the infinite distance case, then this can be further simplified
\be
I_{0,n}(0)=\int_0^1\frac{u^n}{\sqrt{1-u^2}}\dd u
=
\frac{(n-1)!!}{n!!}\times\begin{cases}
\displaystyle  \frac{\pi}{2},&~~n>0\text{ is even},\\
\displaystyle  1,&~~n>0\text{ is odd}.
\end{cases}
\ee

The $I_{n,m,k}$ in Eq. \eqref{eq:iniint} is defined as
\be
I_{n,m,k}=  \int_{1}^{u_i}\dd u G_{n,m,k}(u) \frac{E^{2k} \lb u+1 \rb^{m-k} \kappa^{1-\delta_{m+n,k} }}{\lb E^2-\kappa\rb^{n+m} u^{n+1}(u^2-1)^{m+1/2}} 
\ee
where the first few $G_{n,m,k}$ are given in Eq. \eqref{eq:gfirstfew}. These integrals can be carried out to find $I_{n,m,k}(u_i)$. The first few needed in Eq. \eqref{eq:gfirstfew} are 
\begin{subequations}\label{eq:ifirstfewmaple}
\begin{align}
I_{0,0,0}(u) = & \frac{\pi}{2} - \arctan \lb \frac{1}{\sqrt{u^2-1}} \rb ,\\
I_{0,1,0}(u) = & - \frac{\sqrt{\pi} \kappa \Gamma\lb 2+\frac{3\omega}{2} \rb}{2 \lb 2+3\omega \rb \lb E^2-\kappa \rb \Gamma\lb \frac{3}{2}-\frac{3\omega}{2} \rb} + \frac{u^{-2-3\omega} \kappa\ _2F_1\lb \frac{1}{2},\frac{2+3\omega}{2}; \frac{4+3\omega}{2}; \frac{1}{u^2} \rb }{2 \lb 2+3\omega \rb \lb E^2-\kappa \rb } ,\\
I_{0,1,1}(u) = & - \frac{E^2}{2 \lb E^2-\kappa \rb} \lsb \frac{1}{\sqrt{u^2-1}} - \frac{2\sqrt{\pi}\Gamma\lb 3+ \frac{3\omega}{2} \rb}{\lb 4+3\omega \rb \Gamma\lb \frac{3}{2} + \frac{3\omega}{2} \rb} - \frac{u^{-4+3\omega}\ _2F_1\lb \frac{3}{2}, 2+\frac{3\omega}{2}; 3+\frac{3\omega}{2}; \frac{1}{u^2} \rb}{4+3\omega} \rsb ,\\
I_{1,0,0}(u) = & -\frac{M\kappa \sqrt{u^2-1}}{ (E^2-\kappa)u} ,\\
I_{1,0,1}(u) = & \frac{E^2M (2 u+1) \sqrt{u^2-1}}{ (E^2-\kappa) u (u+1)} .
\end{align}
\end{subequations}

It is also desirable in Eq. \refer{eq:dphirescase2} to have the large $u$ expansion of the above formulas. Carrying out this expansion, we have
\begin{subequations}\label{eq:ifirstfewmapleexpand}
\begin{align}
I_{0,0,0}(u) = & \frac{\pi }{2} - \frac{1}{u} - \frac{1}{6}\frac{1}{u^3} + \mathcal{O}\lb\frac{1}{u^4}\rb , \\
I_{0,1,0}(u) = & \frac{\kappa}{2\lb E^2-\kappa \rb} \lsb - \frac{\sqrt{\pi } \Gamma \lb 2+\frac{3\omega}{2}\rb}{\lb 2+3\omega \rb \Gamma \lb \frac{3}{2} + \frac{3\omega}{2} \rb } + \frac{u^{-2-3\omega}}{ 2+3\omega} + \frac{u^{-4-3\omega} }{2 \lb 4+3\omega \rb } + \frac{u^{-6-3\omega } }{24 \lb 2+\omega \rb } \rsb + \mathcal{O}\lb\frac{1}{u^4}\rb ,\\
I_{0,1,1}(u) = & -\frac{E^2}{E^2-\kappa} \lsb - \frac{\sqrt{\pi } \Gamma \lb 3+\frac{3\omega }{2}\rb}{\lb 4+3\omega \rb \Gamma \lb \frac{3}{2} + \frac{3\omega }{2}\rb } +\frac{1}{2u}+\frac{1}{4 u^3} - \frac{u^{-4-3\omega}}{2 \lb 4+3\omega \rb } - \frac{3u^{-6-3\omega}}{12\lb 2+\omega \rb } \rsb+ \mathcal{O}\lb\frac{1}{u^4}\rb ,\\
I_{1,0,0}(u) = & -\frac{\kappa  M}{E^2-\kappa }+\frac{\kappa  M}{ 2\lb E^2- \kappa \rb}\frac{1}{u^2}+\mathcal{O}\lb\frac{1}{u^4}\rb ,\\
I_{1,0,1}(u) = & \frac{2 E^2 M}{E^2-\kappa }-\frac{E^2 M}{ \lb E^2-\kappa \rb}\frac{1}{u}-\frac{E^2 M}{2\lb E^2-\kappa \rb}\frac{1}{u^3}+\mathcal{O}\lb\frac{1}{u^4}\rb  .
\end{align}
\end{subequations}

Lastly, the $I^{'}_{n,m,k}$ in Eq. \eqref{eq:dtrescase2} is defined as
\be
I^{'}_{n,m,k}=  \int_{1}^{u_i}\dd u G^{'}_{n,m,k}(u) \frac{E^{2k+1} \lb u+1 \rb^{m-k} \kappa^{1-\delta_{m+n,k} }}{\lb E^2-\kappa\rb^{n+m+1/2} u^{n-1}(u^2-1)^{m+1/2}} 
\ee
where the first few $G^{'}_{n,m,k}$ are given in Eq. \eqref{eq:gfirstfewtt}. These integrals can also be carried out to find $I^{'}_{n,m,k}(u_i)$. The first few needed in Eq. \eqref{eq:dtrescase2} are 
\begin{subequations}\label{eq:ifirstfewmaplett}
\begin{align}
I^{'}_{0,0,0}(u) = & \frac{E\sqrt{u^2-1}}{\sqrt{E^2-\kappa}} ,\\
I^{'}_{0,1,0}(u) = & \frac{E\kappa}{2 \lb E^2-\kappa\rb^{3/2}} \lsb 
\frac{6\sqrt{\pi} \Gamma\lb 2+\frac{3\omega}{2} \rb}{\lb 2+3\omega \rb \Gamma\lb \frac{1}{2}+\frac{3\omega}{2} \rb}
- \frac{2\sqrt{\pi} \Gamma\lb 1+\frac{3\omega}{2} \rb}{\omega \Gamma\lb -\frac{1}{2}+\frac{3\omega}{2} \rb} 
+ \frac{u^{-3\omega} \ _2F_1\lb \frac{3}{2},\frac{3\omega}{2}; 1+\frac{3\omega}{2}; \frac{1}{u^2} \rb }{\omega} \right. \nonumber \\
 & \left. - \frac{3u^{-2-3\omega} \ _2F_1\lb \frac{3}{2},1+\frac{3\omega}{2}; 2+\frac{3\omega}{2}; \frac{1}{u^2} \rb }{2+3\omega} \rsb ,\\
I^{'}_{0,1,1}(u) = & - \frac{E^3}{2\lb E^2-\kappa\rb^{3/2}} \lsb 
\frac{1}{\sqrt{u^2-1}}
+\frac{6\sqrt{\pi} \Gamma\lb 2+\frac{3\omega}{2} \rb}{\lb 2+3\omega \rb \Gamma\lb \frac{1}{2}+\frac{3\omega}{2} \rb}
- \frac{4\sqrt{\pi} \Gamma\lb 1+\frac{3\omega}{2} \rb}{3\omega \Gamma\lb -\frac{1}{2}+\frac{3\omega}{2} \rb} 
+ \frac{2u^{-3\omega} \ _2F_1\lb \frac{3}{2},\frac{3\omega}{2}; 1+\frac{3\omega}{2}; \frac{1}{u^2} \rb }{3\omega} \right. \nonumber \\
 & \left. - \frac{3u^{-2-3\omega} \ _2F_1\lb \frac{3}{2},1+\frac{3\omega}{2}; 2+\frac{3\omega}{2}; \frac{1}{u^2} \rb }{2+3\omega} \rsb ,\\
I^{'}_{1,0,0}(u) = & -\frac{3E\kappa M \ln \lb u+\sqrt{u^2-1} \rb}{ (E^2-\kappa)^{3/2}} ,\\
I^{'}_{1,0,1}(u) = & \frac{E^3M }{ (E^2-\kappa)^{3/2}} \lsb \frac{u-1}{\sqrt{u^2-1}} + 2\ln \lb u+\sqrt{u^2-1} \rb \rsb .
\end{align}
\end{subequations}
It is also desirable in Eq. \refer{eq:dtrescase2} to have the large $u$ expansion of the above formulas. Carrying out this expansion, we have
\begin{subequations}\label{eq:ifirstfewmapleexpandtt}
\begin{align}
I^{'}_{0,0,0}(u) = & \frac{E}{\sqrt{E^2-\kappa}} \lb u-\frac{1}{2u}-\frac{1}{8u^3} \rb + \mathcal{O}\lb\frac{1}{u^4}\rb , \\
I^{'}_{0,1,0}(u) = & \frac{E\kappa}{2\lb E^2-\kappa \rb^{3/2}} \lcb 
\frac{6\sqrt{\pi} \Gamma\lb 2+\frac{3\omega}{2} \rb}{\lb 2+3\omega \rb \Gamma\lb \frac{1}{2}+\frac{3\omega}{2} \rb}
- \frac{2\sqrt{\pi} \Gamma\lb 1+\frac{3\omega}{2} \rb}{\omega \Gamma\lb -\frac{1}{2}+\frac{3\omega}{2} \rb} \right. \nonumber \\ 
 & \left. + u^{-3\omega} \lsb \frac{1}{\omega} + \frac{3}{\lb 4+6\omega \rb u^2} + \frac{9}{\lb 32+24\omega \rb u^4} + \frac{5}{16\lb 2+\omega \rb u^6} \rsb \rcb
 + \mathcal{O}\lb\frac{1}{u^4}\rb ,\\
I^{'}_{0,1,1}(u) = & - \frac{E^3}{2\lb E^2-\kappa \rb^{3/2}} \lcb 
\frac{6\sqrt{\pi} \Gamma\lb 2+\frac{3\omega}{2} \rb}{\lb 2+3\omega \rb \Gamma\lb \frac{1}{2}+\frac{3\omega}{2} \rb}
- \frac{4\sqrt{\pi} \Gamma\lb 1+\frac{3\omega}{2} \rb}{3\omega \Gamma\lb -\frac{1}{2}+\frac{3\omega}{2} \rb} 
+ \frac{1}{u} + \frac{1}{2u^3} + \right. \nonumber \\ 
 & \left. u^{-3\omega} \lsb \frac{2}{3\omega} - \frac{3}{4\lb 4+3\omega \rb u^4} - \frac{5}{12\lb 2+\omega \rb u^6} \rsb \rcb
 + \mathcal{O}\lb\frac{1}{u^4}\rb  ,\\
I^{'}_{1,0,0}(u) = & -\frac{3E\kappa M}{\lb E^2-\kappa\rb^{3/2} } \lsb \ln \lb 2u \rb - \frac{1}{4u^2} \rsb + \mathcal{O}\lb\frac{1}{u^4}\rb ,\\
I^{'}_{1,0,1}(u) = & \frac{E^3 M}{\lb E^2-\kappa  \rb^{3/2}} \lsb 2\ln\lb 2u \rb +1 - \frac{1}{u} - \frac{1}{2u^3} \rsb + \mathcal{O}\lb\frac{1}{u^4}\rb  .
\end{align}
\end{subequations}

\section{Deflection in Weyl gravity \label{app:b}}

The Weyl gravity is described by the line element 
\eqref{eq:sssmetric} with metric functions \cite{Edery:1997hu}
\be A(r)=\frac{1}{B(r)}=1-\frac{2M}{r}-\alpha r-kr^2, ~C(r)=r^2. \label{eq:weylmetric}\ee
To show that the deflection angle for null signal in Weyl gravity, when expressed in terms of $r_0$, is equivalent to the deflection angle of the Kiselev BH spacetime with $\omega=-2/3$, we can directly start from Eq. \eqref{eq:deltaphi}. After substituting Eq. \eqref{eq:linr0} and $\kappa=0$, we have 
\be
\Delta \phi
= \lsb \int_{r_0}^{r_s} + \int_{r_0}^{r_d} \rsb  \sqrt{\frac{ABC(r_0)}{C}} \sqrt{\frac{1}{A(r_0) C  - AC(r_0) }}\dd r.
\label{eq:deltaphilsub}
\ee
Then substituting Eq. \eqref{eq:weylmetric} into this, it becomes
\be
\Delta \phi
= \lsb \int_{r_0}^{r_s} + \int_{r_0}^{r_d} \rsb  \frac{r_0}{r} \lsb \lb 1-\frac{2M}{r_0}-\alpha r_0\rb r^2 -\lb 1-\frac{2M}{r}-\alpha r\rb r_0^2 \rsb^{-1/2}\dd r.
\label{eq:deltaphiwl}
\ee
Clearly, the $k$ parameter in metric \eqref{eq:weylmetric} does not appear in $\Delta \phi$, and this is exact the deflection angle of null signals in the Kiselev BH spacetime  \eqref{eq:qmetric} with $\omega=-2/3$.

The perturbative deflection angle of null signals in Weyl gravity, or equivalently in the Kiselev BH spacetime with $\omega=-2/3$, then can be directly obtained from Eq.  \eqref{eq:dphirescase2} by substituting $\kappa=0$. Its expansion to the first few non-trivial orders of $(M/r_0),~(r_0/r_{s,d})$ and $\alpha$ is 
\begin{align} \Delta\phi=& 
\pi+\frac{4 M}{r_0}-
   \left(\frac{r_0}{r_s}+\frac{r_0}{r_d}\right)
+\left(\frac{15 \pi }{4}-4\right) \frac{M^2}{r_0^2}
- \left(\frac{M}{r_s}+\frac{M}{r_d}\right)
-\frac{3}{2}\frac{M}{r_0}\lb \frac{M}{r_s}+\frac{M}{r_d} \rb
-\frac{1}{6}\lb \frac{r_0^3}{r_s^3}+\frac{r_0^3}{r_d^3} \rb \nonumber \\
& +\alpha r_0 \left[1+\left(\frac{3 \pi }{2}-1\right) \frac{M}{r_0}- \left(\frac{r_0}{2 r_d}+\frac{r_0}{2 r_s}\right) 
+ \lb \frac{47}{2}-\frac{15\pi}{4} \rb \frac{M^2}{r_0^2} 
-\frac{3}{2} \lb \frac{M}{r_s} + \frac{M}{r_d} \rb 
+\frac{1}{4} \lb \frac{r_0^2}{r_s^2}+\frac{r_0^2}{r_d^2} \rb \right] \nonumber \\
& + \alpha^2 r_0^2 \left[ \frac{1}{2} + \frac{3M}{r_0} -\frac{3}{8} \lb \frac{r_0}{r_s}+\frac{r_0}{r_d} \rb \right] \label{eq:defweyl}
. \end{align}
Its infinite $r_s,~r_d$ limit  can be easily obtained. 

\section{$\Delta\phi$ and $\Delta t$ in terms of $b$ for case $-1\leq\omega < -1/3$ \label{app:c}}

The deflection angle in Eq. \eqref{eq:c2daser} for the case $-1\leq\omega<-1/3$ was given as a series involving the closet radius $r_0$, which can be determined by combining Eqs. \eqref{eq:leandbv} and \eqref{eq:pfuncdef}. On the other hand, deflection angles are more often expressed as a series of the impact parameter $b$, which has a one-to-one correspondence with $r_0$. The impact parameter $b$ also has a simple and intuitive geometrical interpretation as the distance from the lens center to the asymptotic straight line in an asymptotically flat spacetime. Although in general in the asymptotically non-flat cases, this interpretation is not strictly valid anymore, nevertheless mathematically we can still define an effective impact parameter using the same Eq. \eqref{eq:leandbv} and express $\Delta \phi$ in terms of $b$, as done in Refs. \cite{Rindler:2007zz,Takizawa:2020egm}. This effective $b$ is still able to characterize the scale of minimal distance of the trajectory to the lens center in the small $\alpha$ cosmological constant limit. 
Here we present $\Delta\phi$, as well as the total travel time $\Delta t$, in terms of $b$ for potential future reference. 

Using Eq. \refer{eq:pfuncdef}, in the small $\alpha$ and $r_0\sim b\gg M$ limits, we can write  $r_0$ as a series of $b$ as
\begin{align}
r_0 =& b - \frac{E^2M}{E^2-\kappa} -\frac{E^2\lb 3E^2-4\kappa \rb M^2}{2\lb E^2-\kappa \rb^2 b} - \lcb \frac{E^2 b}{2\lb E^2-\kappa \rb } +\frac{E^2\lsb 3\lb 1+\omega\rb E^2-4\kappa \rsb M}{2\lb E^2-\kappa\rb^2} \rcb b^{-1-3\omega}\alpha \nonumber\\
 & - \frac{3\omega E^4}{4\lb E^2-\kappa\rb^2} b^{-1-6\omega} \alpha^2 + \mathcal{O} \lb \frac{1}{b^2},\frac{\alpha}{b^{2+3\omega}} \rb.
\end{align}
Substituting this into the deflection angle \refer{eq:c2daser} and total flight time \refer{eq:c2ttser}, to the order $(b/r_i)^1,~(M/b)^1$ and $\alpha^1$, they are transformed to
\begin{align}
    \Delta \phi = & \sum_{i=s,d} \lsb \frac{\pi}{2} - \frac{b}{r_i} + \frac{\lb 2E^2-\kappa \rb M}{\lb E^2-\kappa \rb b} + \lcb \frac{\kappa}{2\lb E^2-\kappa \rb \lb 2+3\omega \rb} \lsb  \lb\frac{ b}{r_i}\rb^{2+3\omega}  -\frac{\sqrt{\pi } \Gamma \lb 2+\frac{3\omega }{2}\rb  }{ \Gamma \lb \frac{3}{2}+\frac{3\omega }{2} \rb } \rsb \right.\right. \nonumber \\
& \left.\left. +\frac{E^2}{E^2-\kappa}  \frac{\sqrt{\pi } \Gamma \lb 3+\frac{3\omega}{2}\rb  }{\lb 4+3\omega  \rb \Gamma \lb \frac{3}{2}+\frac{3\omega }{2}\rb }  \rcb \frac{\alpha}{b^{1+3\omega}} \rsb  + \mathcal{O} \lb \varepsilon^2 \rb ,\label{eq:dphicase2inb}\\
\Delta t = & \sum_{i=s,d} \frac{E}{\sqrt{E^2-\kappa}} \lcb r_i - \frac{b^2}{2r_i} + M \lb \frac{ 2E^2-3\kappa }{E^2-\kappa}\ln \frac{2r_i}{b} + \frac{E^2}{E^2-\kappa}  \rb - \frac{\alpha}{b^{3\omega}} 
  \lsb \frac{3\sqrt{\pi} \Gamma\lb 2+\frac{3\omega}{2} \rb}{\lb 2+3\omega \rb \Gamma\lb \frac{1}{2}+\frac{3\omega}{2} \rb} \right. \right.   \nonumber \\
 & \left.\left.  - \frac{\lb 2E^2-3\kappa\rb \sqrt{\pi} \Gamma\lb 1+\frac{\omega}{2} \rb}{3\lb E^2-\kappa\rb \omega \Gamma\lb -\frac{1}{2}+\frac{3\omega}{2} \rb} 
 + \frac{2E^2-3\kappa}{6\lb E^2-\kappa\rb\omega} \lb \frac{r_i}{b}\rb^{-3\omega}\rsb\rcb + \mathcal{O} \lb \varepsilon^2 \rb. \label{eq:dtcase2inb}
\end{align}
Note that in Eq. \eqref{eq:dphicase2inb}, the effect of finite distance of the source and detector are taken into account too. 

\end{widetext}

\end{document}